\documentclass{osa-article-noc}

\newcommand{\fir}[1]{Fig.~\ref{#1}}
\newcommand{\lfir}[1]{Figure \ref{#1}}
\newcommand{\secr}[1]{Sec.~\ref{#1}}
\newcommand{\eqr}[1]{Eq.~(\ref{#1})}

\usepackage{physics}
\usepackage[per-mode = symbol]{siunitx}
\usepackage{booktabs}
\usepackage{textcomp}
\usepackage{tikz}
\usepackage{multirow}
\usepackage{makecell}

\journal{oe}

\begin{document}

\title{A high-flux, adjustable, compact cold-atom source}

\author{Sean Ravenhall,\authormark{1,2} Benjamin Yuen,\authormark{3} and Chris Foot\authormark{1,*}}

\address{
		\authormark{1}Clarendon Laboratory, University of Oxford, Parks Road, Oxford, OX1 3PU, UK\\
		\authormark{2}PA Consulting, Melbourn, SG8 6DP, UK\\
		\authormark{3}School of Physics and Astronomy, University of Birmingham, Birmingham, B15 2TT, UK
	}

\email{\authormark{*}christopher.foot@physics.ox.ac.uk}

\begin{abstract}
	Magneto-optical traps (MOTs) are widely used for laser cooling of atoms. We have developed a high-flux compact cold-atom source based on a pyramid MOT with a unique adjustable aperture that is highly suitable for portable quantum technology devices, including space-based experiments. The adjustability enabled an investigation into the previously unexplored impact of aperture size on the atomic flux, and optimisation of the aperture size allowed us to demonstrate a higher flux than any reported cold-atom sources that use a pyramid, LVIS, 3D-MOT or grating MOT. We achieved $2.0(1) \times 10^{10}$\,atoms/s of $^{87}$Rb with a mean velocity of 32(1)\,m/s, FWHM of 27.6(9)\,m/s and divergence of 58(3)\,mrad. Halving the total optical power to 195\,mW caused only a 26\% reduction of the flux, and a 33\% decrease in mean velocity. Methods to further decrease the velocity as required have been identified. The low power consumption and small size make this design suitable for a wide range of cold-atom technologies.
\end{abstract}

\section{Introduction}

Cold-atom quantum technology is increasingly being reimagined from research apparatus into practical devices operating in real-world environments. These precision measuring instruments provide significant improvement in performance over classical techniques in applications including timekeeping \cite{Schioppo2017Ultrastable, Poli2014Transportable, Ludlow2015Optical, Elvin2019Cold, Bowden2019Pyramid}, gravimetry \cite{Grotti2018Geodesy, Bodart2010Cold, Bidel2013Compact, Gillot2016LNE, Menoret2018Gravity, Bidel2018Absolute, Freier2016Mobile, Wu2019Gravity, Wang2018Shift, Mazon2019Portable} and inertial sensing for navigation \cite{Geiger2011Detecting, Wu2017Multiaxis, Battelier2016Development, Cheiney2018Navigation}. Compact cold-atom devices are also being developed for space-based experiments for gravity mapping, navigation and communications, as well as fundamental physics research in general relativity, dark matter and gravitational waves \cite{Becker2018Space, Elliott2018NASA, Trimeche2019Concept, Chiow2015Laser, Liu2018In-Orbit, Hogan2011Atomic, Gehler2013ESA, Williams2016Quantum, Hogan2016Atom, Tino2019SAGE, Schuldt2015Design, Loriani2019Atomic, Dimopoulos2009Gravitational}.

A key component for such devices is a cold-atom source, which generates a beam of laser-cooled atoms directed towards a region where precision measurements are carried out. Although the techniques used for the measurements vary widely depending on the specific application, a source is common to the vast majority of cold-atom systems. This means that a compact and robust source is an essential component for enabling practical quantum devices. Such a source is required to produce a high flux of atoms that have a trappable velocity (typically 20-50\,m/s when loading a magneto-optical trap (MOT) \cite{Foot2005Atomic}). A higher flux improves the repetition rate, duty cycle, signal-to-noise and measurement bandwidth. The atomic beam should have a low divergence and a low background of thermal atoms that have not been cooled, to facilitate subsequent trapping, while efficiently utilising the available optical power.

A MOT source formed by mirrors arranged in a pyramid configuration has many advantages. Such a pyramid MOT source achieves compactness and robustness by reflecting a single large-diameter laser beam using an in-vacuum pyramidal reflector to produce the optical fields required for laser cooling \cite{Lee1996Single}. An aperture at the apex of the pyramid creates a region without retroreflected light, and the resulting intensity imbalance causes atoms that enter this `exit' channel to be pushed towards the aperture and out of the cooling region, forming an atomic beam \cite{Arlt1998Pyramidal, Williamson1998Magneto} instead of a trapped cloud (see \fir{f:PyramidMOT}). As well as simplicity, the pyramid source approach has the advantages of: balanced intensities at all positions; all phase and intensity noise is common-mode between beam pairs \cite{Hinton2017Portable}; and it only requires optical access for a single incoming laser beam, because the light is retroreflected along the axis of incident (single) laser beam.

\begin{figure}[t]
	\centering
	\resizebox{0.80\linewidth}{!}{
		\begin{tikzpicture}
		\begin{scope}
		\node[anchor=south west,inner sep=0] (image) at (0,0) {
			\resizebox{0.9\textwidth}{!}{
				\includegraphics[trim={0 14 170 0}, clip]{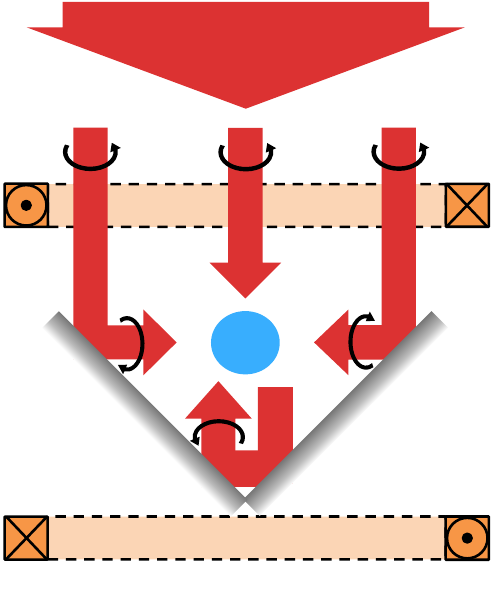}
				\quad\quad\quad\quad\quad\quad
				\includegraphics[trim={170 14 0 0}, clip]{Figures/PyramidMOT}
			}
		};
		\end{scope}
		\end{tikzpicture}
	}
	\caption{A pyramid MOT (left) cools and traps atoms (blue circle) by illuminating a (usually square-based) pyramid assembly of mirrors with circularly polarised light, in the presence of a spherical quadrupole magnetic field (generated by an anti-Helmholtz coil pair, orange). In a pyramid MOT source (right), an aperture at the pyramid apex leads to an absence of retroreflected light along the axis (the `exit' region) where the intensity imbalance causes atoms to be pushed out through the aperture, forming an atomic beam.}
	\label{f:PyramidMOT}
\end{figure}

Alternative approaches to cold-atom sources of alkali metals include Zeeman slowers, 2D \& 2D$^+$ MOT sources, low-velocity intense sources (LVISs), and grating MOT sources. Table~\ref{t:Sources} summarises the highest-flux examples reported for each type of source, along with some other high-flux examples for additional comparison. The highest fluxes are provided by Zeeman slowers, at the cost of larger and more complicated apparatus. LVIS and 2D \& 2D$^+$ MOT sources are popular in research laboratories because they achieve high fluxes while requiring few or no in-vacuum optics. This makes them simpler to construct than pyramid sources, but less compact and more susceptible to mechanical misalignments. Grating MOTs achieve a highly compact optical arrangement by diffracting a single laser beam from a planar grating reflection hologram, however the flux of atoms is low and the use of diffraction gratings causes significant optical power loss into diffraction orders and polarisation states other than those required for laser cooling.

\begin{table}[t]
	\sisetup{
	}
	\centering
	\def\arraystretch{1.3}
	\resizebox{1.0\textwidth}{!}{
		\begin{tabular}{llc
				S[table-format = 1.4e1,
				table-sign-exponent
				]ccccccc}
			\toprule
			\multirow{2}{*}[.3ex]{Source Type} & \multirow{2}{*}[.3ex]{Institution} & \multirow{2}{*}[.3ex]{Year}  & {Flux} & \multicolumn{2}{c}{Velocity (m/s)} & Divergence & Power & Intensity & \multirow{2}{*}[.3ex]{Ref.}         \\[-.8ex]
			                       &                  &      & {(atoms/s)} & Mean  &   FWHM   &  (mrad)  & (mW) & ($I_\textnormal{sat}$) &                                 \\ \midrule
			\textbf{Zeeman Slower} & Havard           & 2005 &    1.7e12   &  45   &    8.6   &    43    &  400 &          --            &       \cite{Slowe2005High}      \\ % $^{87}$Rb
			                       & AOSense*         & 2018 &      1e11   &  40   &    --    &    25    &  --  &          --            &          \cite{AOSense}         \\[1ex] % Sr/Ca/Yb
			\textbf{Pure 2D-MOT}   & FOM, Amsterdam   & 1998 &      6e9    &  12   &    15    &    46    & 32.8 &          1.9           &     \cite{Dieckmann1998Two}     \\ % $^{87}$Rb
			                       & Stuttgart        & 2002 &      6e10   &  50   &    75    &    32    &  640 &         10.2           &    \cite{Schoser2002Intense}    \\[1ex] % $^{87}$Rb
			\textbf{2D-MOT}        & LKB, Paris       & 2002 &      1e9    & 0.3-3 & 0.07-0.7 & 400-100  &  120 &          2.5           &      \cite{Cren2002Loading}     \\ % $^{87}$Rb
			                       & JPL, Pasadena*   & 2012 &      8e9    &  --   &    --    &    --    &  20  &          2.1           &    \cite{Kellogg2012Compact}    \\ % $^{133}$Cs
			                       & KRISS, Daejeon   & 2012 &    4.5e10   &  39   &    2     &    12    &  60  &          2.7           &       \cite{Park2012Cold}       \\[1ex] % $^{87}$Rb
			\textbf{2D$^+$-MOT}    & FOM, Amsterdam   & 1998 &      9e9    &   8   &   3.3    &    43    & 32.8 &          1.7           &     \cite{Dieckmann1998Two}     \\ % $^{87}$Rb
			                       & TIFR, Mumbai     & 2006 &      2e10   &  17   &    5     &    26    &  55  &          2.4           & \cite{Chaudhuri2006Realization} \\ % $^{87}$Rb
			                       & Hannover         & 2011 &    8.4e10   & 17.7  &    9     &    --    &  309 &          4.4           & \cite{Jollenbeck2011Hexapole, Jollenbeck2012Eine} \\ % $^{87}$Rb
			                       & ColdQuanta       & 2018 &      1e9    &  --   &    --    &    --    &  --  &          --            &        \cite{ColdQuanta}        \\[1ex] % Rb/Cs
			\textbf{3D-MOT}        & LKB, Paris       & 2001 &    1.3e8    &  14   &    21    &    12    &  30  &          3.0           &     \cite{Wohlleben2001Atom}    \\[1ex] % $^{87}$Rb
			\textbf{LVIS}          & NIST, Boulder*   & 2005 &      1e10   &  11   &    --    &    --    &  100 &          9.1           &     \cite{Donley2005Optical}    \\ % $^{133}$Cs
			                       & NPL, London      & 2005 &      8e9    &  8.5  &    4     &    44    &  58  &          1.4           &  \cite{Ovchinnikov2005Compact}  \\[1ex] % $^{87}$Rb
			\textbf{Pyramid MOT}   & Oxford*          & 1998 &    1.1e9    &  12   &   2-3    &    40    &  86  &          5.0           &     \cite{Arlt1998Pyramidal}    \\ % $^{133}$Cs
			                       & Pisa*            & 2001 &      4e9    & 8-12  &   1.5    & 26$\pm$1 &  100 &          9.1           &     \cite{Camposeo2001Cold}     \\ % $^{133}$Cs
			                       & JPL, Pasadena*   & 2003 &    2.2e9    &  15   &    --    &   11-16  &  600 &          2.7           &    \cite{Kohel2003Generation}   \\ % $^{133}$Cs
			                       & JPL, Pasadena*   & 2004 &      2e8    &  --   &    --    &    15    &  175 &          0.9           &      \cite{Lundblad2004Two}     \\ % $^{133}$Cs
			                       & Durham           & 2008 & 1.8(3)e7    &  --   &    --    &    --    &  50  &          2.5           &    \cite{Harris2008Magnetic}    \\ % $^{87}$Rb
			                       & Oxford    & 2020 & \multicolumn{1}{c}{$2.0(1) \times 10^{10}$} & 32(1) & 27.6(9) & 58(3) & 390 & 6.0 & \multirow{2}{*}{\makecell[cc]{This\\work}}	\\ % $^{87}$Rb
			                       & Oxford (50\% power) & 2020 & \multicolumn{1}{c}{$1.6(1) \times 10^{10}$} & 22.6(7) & 18.6(6) & 67(4) & 195 & 3.0 &          		   \\[1ex] % $^{87}$Rb
			\textbf{Grating MOT}   & AFRL, New Mexico & 2017 &      4e8    &16.5(9)&   9(7)   &    --    & 51.5 &          6.6           & \cite{Imhof2017Two-dimensional} \\ % $^{87}$Rb
			\bottomrule
		\end{tabular}
	}
	\caption{A summary of cold atom sources from the literature, sorted by source type, with (where available) the: total flux; mean and FWHM of the axial velocity distribution; full-angle divergence, using the FWHM; total optical cooling power used (excluding repumping power); and the average cooling intensity of the beams, averaged over the $1/e^2$ diameter, using $I_\textnormal{sat}=1.67$\,mW/cm$^2$ for $^{87}$Rb \cite{Steck2015Rubidium} and $I_\textnormal{sat}=1.10$\,mW/cm$^2$ for $^{133}$Cs \cite{Steck2019Cesium}. All sources used $^{87}$Rb except for those marked * that used $^{133}$Cs (which are mainly included where they are the highest recorded flux of that source type). Note that \cite{AOSense} \& \cite{ColdQuanta} are commercially available devices.
	}
	\label{t:Sources}
\end{table}

We report here a new design of pyramid source that is notable for two main reasons. Firstly, it produces the highest reported flux from a pyramid source in the literature; only three non-Zeeman slower sources are known to produce higher fluxes \cite{Jollenbeck2011Hexapole, Park2012Cold, Schoser2002Intense}, and they are all significantly larger with four independent beams, making them unsuitable for compact devices. Even operating at 50\% intensity our source produces the highest flux of any reported pyramid, LVIS, 3D-MOT or grating-MOT source. Secondly, it has the unique feature of an adjustable aperture size, which we used to investigate optimisation of the source. To our knowledge no investigation of the effect on the flux of varying the aperture size has been performed in the literature. In the following, we first review designs of pyramids in laser cooling, before detailing our new design and the method used to characterise its performance. Finally, we present and analyse the results of this investigation, including the ultimate performance of the source, how it compares to other reported sources, and our findings on the influence of aperture size on the atomic beam.

\section{Pyramid Source Design}

Previous designs of pyramid MOTs have typically used square-based pyramids formed of glass prisms \cite{Arlt1998Pyramidal, Camposeo2001Cold, Harris2008Magnetic, Tierney2009Magnetic} or triangular mirrors \cite{Lee1996Single, Williamson1998Magneto, Vangeleyn2009Single, Li2008Manipulation, Xu2008Realization}, sometimes with a flat retroreflector at the apex of the pyramid \cite{Kohel2003Generation, Lundblad2004Two, Hinton2017Portable}. An aperture at the pyramid apex turns the MOT into a MOT source (see \fir{f:PyramidMOT}) and can be formed by either a gap between the mirrors or a hole drilled through a retroreflector. In all examples that we found in the literature, the aperture size is fixed by the mirror geometry or the hole in the retroreflector, which has prevented experimental investigation into the impact of the aperture size on the atomic beam before now.

Our novel design of pyramid source has the unique feature of an adjustable aperture, as illustrated in \fir{f:Pyramid}. Four identical reflectors are located in straight grooves on a circular `rotator' plate; the grooves confine each reflector to slide along a single direction. The rotator plate sits on a circular `base' plate with similar but curved slots. Each reflector houses a bolt that passes through a slot in both the rotator and base plate, and which can be tightened to secure the reflector in place. As a result, rotating the plates relative to one another causes synchronised movement of the reflectors. This permits the aperture size to be adjusted while maintaining the correct angle of the reflectors and avoiding touching or scratching. Examples of the pyramid set to different aperture sizes are shown in \fir{f:Pyramid}(b).

\begin{figure}[t]
	\centering
	\resizebox{0.9\linewidth}{!}{
		\begin{tikzpicture}
			\begin{scope}
				\node[anchor=south west,inner sep=0] (image) at (0,0) {
					\resizebox{0.9\textwidth}{!}{
						\resizebox{0.55\linewidth}{!}{
							\includegraphics[trim=5 2 5 2, width=0.65\linewidth]{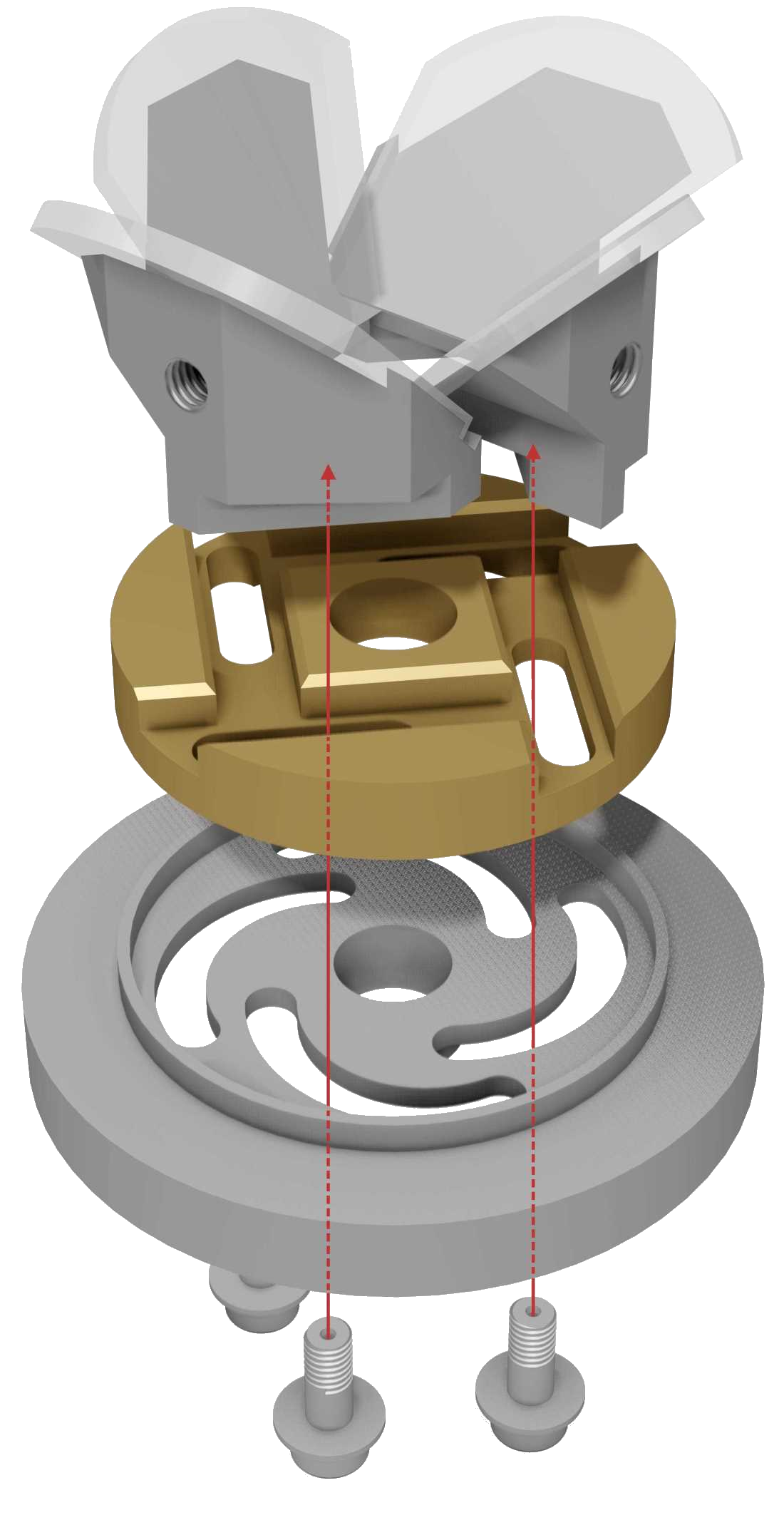}
						}
						\hspace{0.15\linewidth}
						\resizebox{0.23\linewidth}{!}{
							\rotatebox[origin=bl]{90}{
								\hspace{1.945\linewidth}
								\includegraphics[angle=0,origin=c,width=\linewidth]{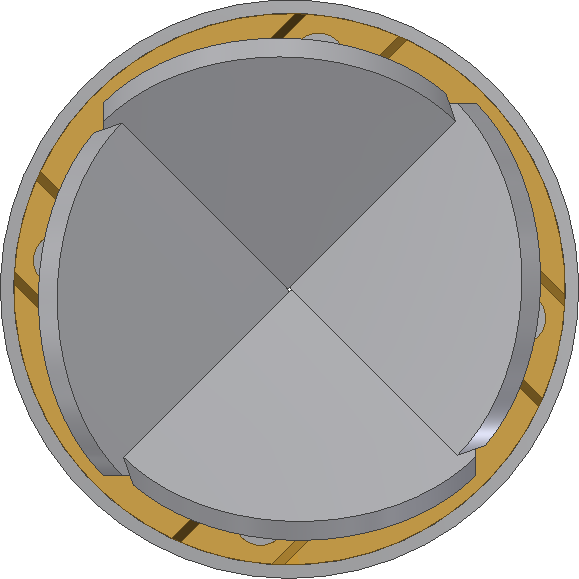}
								\quad
								\includegraphics[angle=0,origin=c,width=\linewidth]{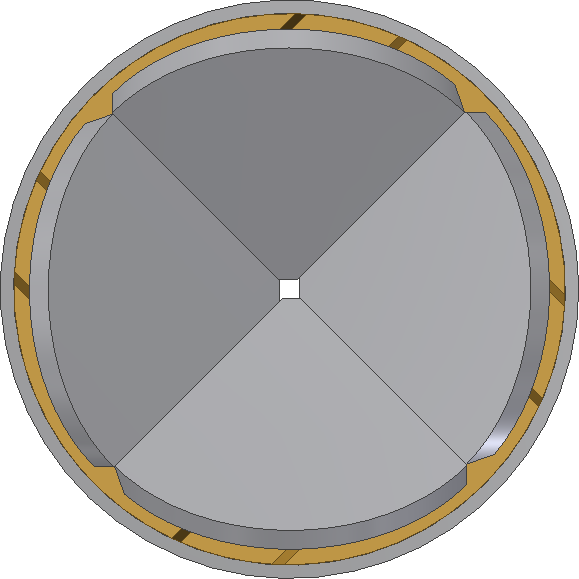}
								\quad
								\includegraphics[angle=0,origin=c,width=\linewidth]{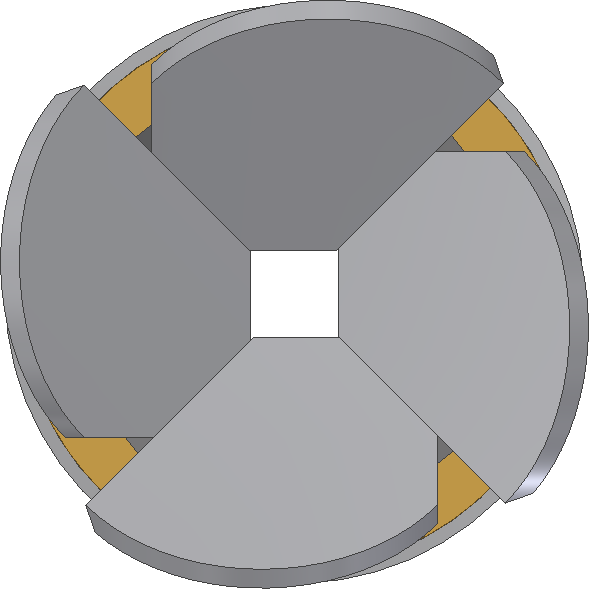}
							}
						}
						\hspace{-0.357\linewidth}
						\includegraphics[trim=0 0 80 0,width=0.4\linewidth]{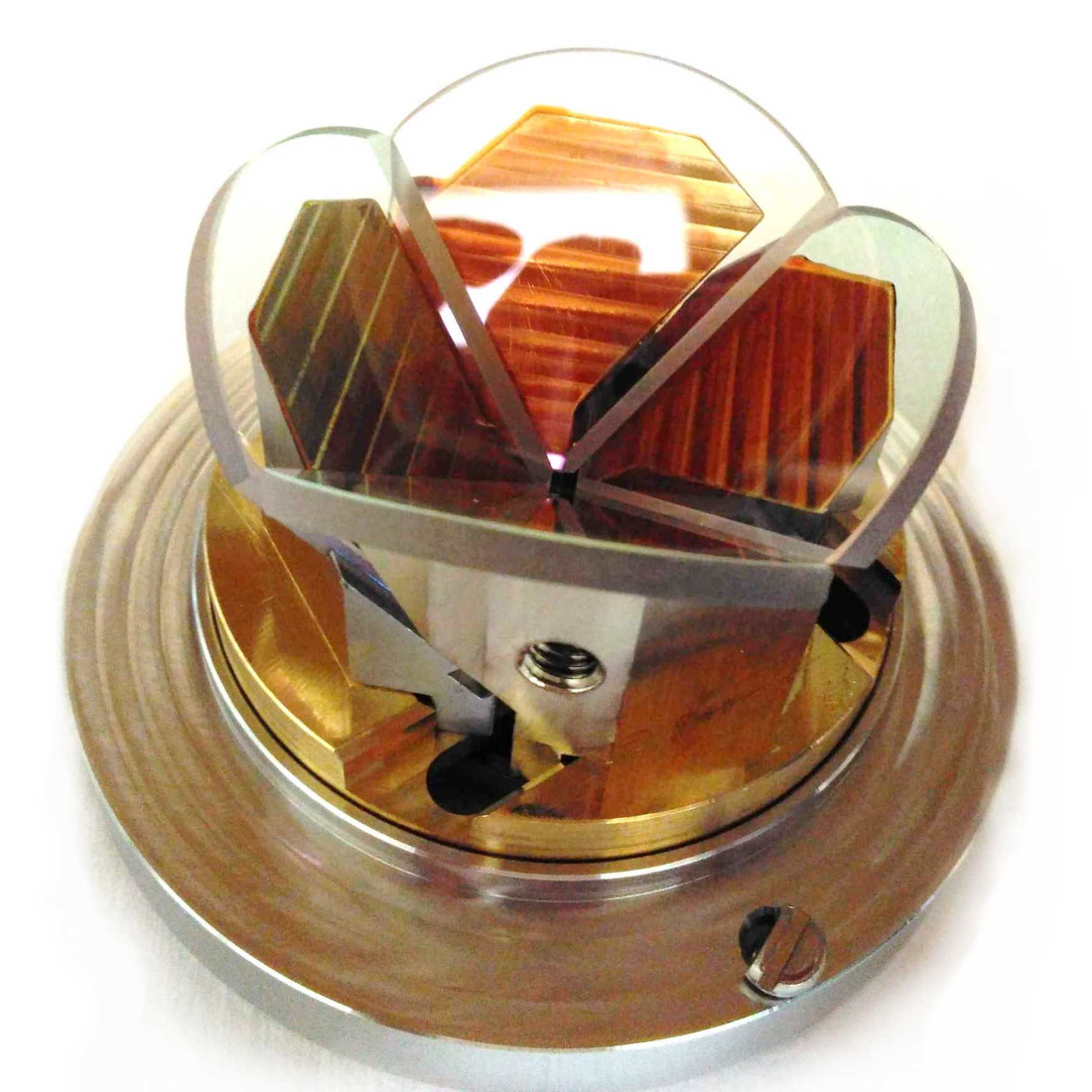}
						\quad
					}
				};
			\end{scope}
			\node [anchor=west] (label1) at (0,12.8) {(a)};
			\node [anchor=west] (label2) at (7.5,12.8) {(b)};
			\node [anchor=west] (label3) at (7.3,4.7) {(c)};
			\definecolor{mygrey}{rgb}{0.42, 0.42, 0.42}
			\node [anchor=west, mygrey] (scalelabel) at (7.1,0.45) {\footnotesize 20\,mm};
			\draw [anchor=west, very thick, mygrey] (7.2,0.2) -- (8.2,0.2);
			\draw [anchor=west, very thick, mygrey] (7.2,0.12) -- (7.2,0.28);
			\draw [anchor=west, very thick, mygrey] (8.2,0.12) -- (8.2,0.28);
		\end{tikzpicture}
	}
	\caption{Adjustable-aperture pyramid design (using polished glass mirror substrates): (a) exploded CAD render; (b) on-axis views for various aperture sizes; (c) the fabricated prototype. Four reflectors (each formed of a multilayer dielectric-coated glass plate glued to a metal mount) are secured by bolts that pass through a `base' plate (grey) and a `rotator' plate (gold). Both plates have slots that the bolts pass through, and the red lines show the path of two of the four bolts. Rotation of the plates relative to each other controls the size of the aperture size at the pyramid apex. Components are made from 316L stainless steel except for the BK7 glass mirrors and the rotator plate, which is aluminium bronze for its self-lubricating properties. More details available in the Supplementary Material.}
	\label{f:Pyramid}
\end{figure}
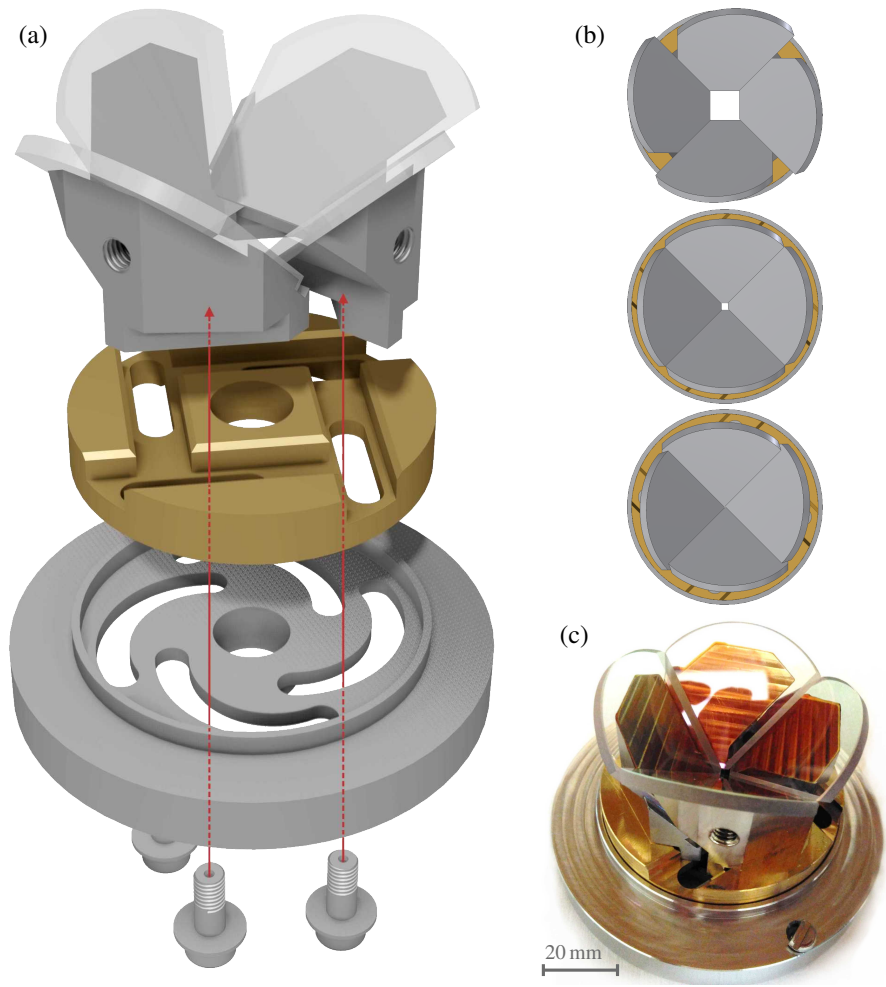

The four identical reflectors are shaped to present a circular profile to the incident laser beam, whereas the majority of pyramids in the literature present a square profile \cite{Camposeo2001Cold, Harris2008Realisation, McCarron2011Quantum, Tierney2009Magnetic, Harsono2006Dipole, Sheard2010Magnetic, Hinton2017Portable}. Illuminating a square-profile pyramid with a circular laser beam is inefficient either in the use of space (not fully illuminating the whole pyramid) or wasting optical power (light missing the pyramid). The reflectors overlap each other to facilitate the adjustable-aperture feature, and the overlapping also minimises gaps between mirrors. This design, resembling the petals of a flower, also reduces the scattering of light from the mirror edges, which can go in all directions and disrupt the laser cooling mechanisms, thereby reducing performance \cite{Arlt1998Pyramidal, Kohel2003Generation, Lu1996Low, Lee1996Single} (in particular a roll off in the flux with increasing intensity).

Two types of reflector were fabricated and tested: polished glass plates secured to 316L stainless steel mounts using EPO-TEK 353ND epoxy (see \fir{f:Pyramid}c); and stainless steel polished to $\lambda/2$ flatness (from LBP Optics, UK), which avoided the need for epoxy. In both cases a multi-layer dielectric coating was applied for high reflectivity at 45\textdegree\ AOI and close to 180\textdegree\ phase shift between the s- and p- components (to produce the correct polarisations for laser cooling). The measured reflectivity of the coating was 94-97\% (these are usable values although less than expectations because of manufacturing issues described more in \secr{s:Apparatus}) and the phase shift was within 18-20\textdegree\ of the target 180\textdegree, which is sufficient for good laser cooling \cite{Arlt1998Pyramidal, Lu1996Low}. Prior to coating, the polished stainless steel mirrors had 54\% reflectivity and a 194\textdegree\ phase shift. The coatings withstood baking at 200\textcelsius\ without noticeable performance degradation.

The pyramid was designed to fit inside a standard DN63CF vacuum tube (typical inner diameter 60\,-\,62\,mm), resulting in an outer pyramid diameter of 59\,mm (with the aperture closed), although the design is scalable to any size. The adjustable-aperture mechanism also aided assembly, since the reflectors could be positioned individually without overlapping, then the mechanism adjusted to overlap them without colliding or scratching.

\section{Apparatus}\label{s:Apparatus}

We operated and characterised the new design of pyramid source with up to 390\,mW of collimated laser light at the pyramid, generated by a titanium-doped sapphire laser (M Squared SolsTiS), although a high flux could be obtained using only half the maximum available power (see \secr{s:MaxFlux}). The usable power was limited by the damage threshold of the optical fibre that transported light to the pyramid. In addition to laser light that is near resonance with the strong transition used for laser cooling of the atoms, the operation of a MOT for alkali atoms also requires light that is on resonance with a transition from the lower hyperfine level of the atom's ground electronic configuration. This so-called repumping light was provided by passing the light through an electro-optic modulator (EOM; Qubig Rb87-6.6G prototype) to generate sidebands; only the lower-frequency sideband is useful. Each first-order sideband had 6\% of the power the carrier (cooling) frequency, and higher order sidebands were negligible. A larger beam diameter than pyramid diameter was chosen (66.7\,mm $1/e^2$, compared to the 55\,mm inner pyramid diameter when the aperture was closed) to provide a more uniform intensity across the pyramid. The average cooling intensity over the $1/e^2$ diameter was 6.0\,$I_\textnormal{sat}$, where $I_\textnormal{sat} = \SI{1.67}{\milli \watt \per \centi \metre \squared}$ is the saturation intensity for the $D_2$ cycling transition ($5^2S_{1/2} \rightarrow 5^2P_{3/2}$) in Rb$^{87}$ for $\sigma^{\pm}$-polarised light \cite{Steck2015Rubidium}. %Specifically, $I_\textnormal{sat}$ is defined as
%\begin{equation}\label{e:Isat}
%	I_\textnormal{sat} = \frac{c \epsilon_0\Gamma^2 \, \hbar^2}{4 \abs{\hat{\boldsymbol{e}} \cdot \boldsymbol{d}}^2}
%\end{equation}
%where $c$ is the speed of light, $\epsilon_0$ is the vacuum permittivity of space, $\Gamma$ is the spontaneous decay rate of the excited state, $\hbar$ is the reduced Planck constant, $\hat{\boldsymbol{e}}$ is the unit polarization vector of the light field, and $\boldsymbol{d}$ is the atomic dipole moment of the transition. Well done for reading this!

Throughout the characterisation, the cooling light detuning was $-2.0\,\Gamma$ and the magnetic quadrupole gradient was \SI{18.2}{G \per \centi \metre} because these were observed to be optimal (the gradient could be reduced to \SI{16}{G \per \centi \metre} before the flux was impacted). All intensities reported are the average cooling intensity over the $1/e^2$ diameter of the Gaussian beams, and the default intensity when not otherwise mentioned was 6.0\,$I_\textnormal{sat}$, since this was the maximum available. All errors are standard errors.

Detection of the atomic flux was performed using absorption spectroscopy (described in \secr{s:Measurement}) with a 1.5\,mm $1/e^2$ diameter collimated probe beam, which was directed transversely through the atomic flux and detected on a photodiode. The probe light was circularly polarised and a weak uniform magnetic field was applied parallel to the probe beam to achieve maximum absorption. The power in the probe beam was 30.0\,\textmu W (1.0\,$I_\textnormal{sat}$) and actively stabilised to 1\%. The Rb$^{87}$ vapour was generated and controlled using dispensers (SAES Getters) connected to high-current feedthroughs in the ultra-high vacuum system. More detailed technical information can be found in Ref. \cite{Ravenhall2018Compact}.

The majority of the data presented in \secr{s:Results} was taken using a pyramid with glass mirrors, however the dielectric coating had abnormally high scatter (because of poor vacuum quality during the deposition of the dielectric coating) which limited the maximum achievable flux of cold atoms. To better determine the maximum flux, a second design of pyramid was produced that had a higher quality dielectric coating on a polished metal substrate, and this gave the peak fluxes reported in the following. The reflectivity was still only 94-95\%, which is below the 99\% expected from multilayer dielectric mirrors, but is sufficient for MOT operation. The two pyramids were identical in geometry, and only the coatings and substrate materials were different.

\section{Measurement Technique}\label{s:Measurement}

The important characteristics of an atomic beam from a cold-atom source are: the total flux of atoms, the atomic velocity distribution, and the beam divergence. These were determined by combining transit time measurements of the atomic velocities with absorption measurements of the atomic density, as follows.

\begin{figure}
	\centering
	\resizebox{0.9\textwidth}{!}{
		\begin{tikzpicture}
			\begin{scope}
			\node[anchor=south west,inner sep=0] (image) at (0,0) {
				\resizebox{0.9\textwidth}{!}{
					\includegraphics[clip]{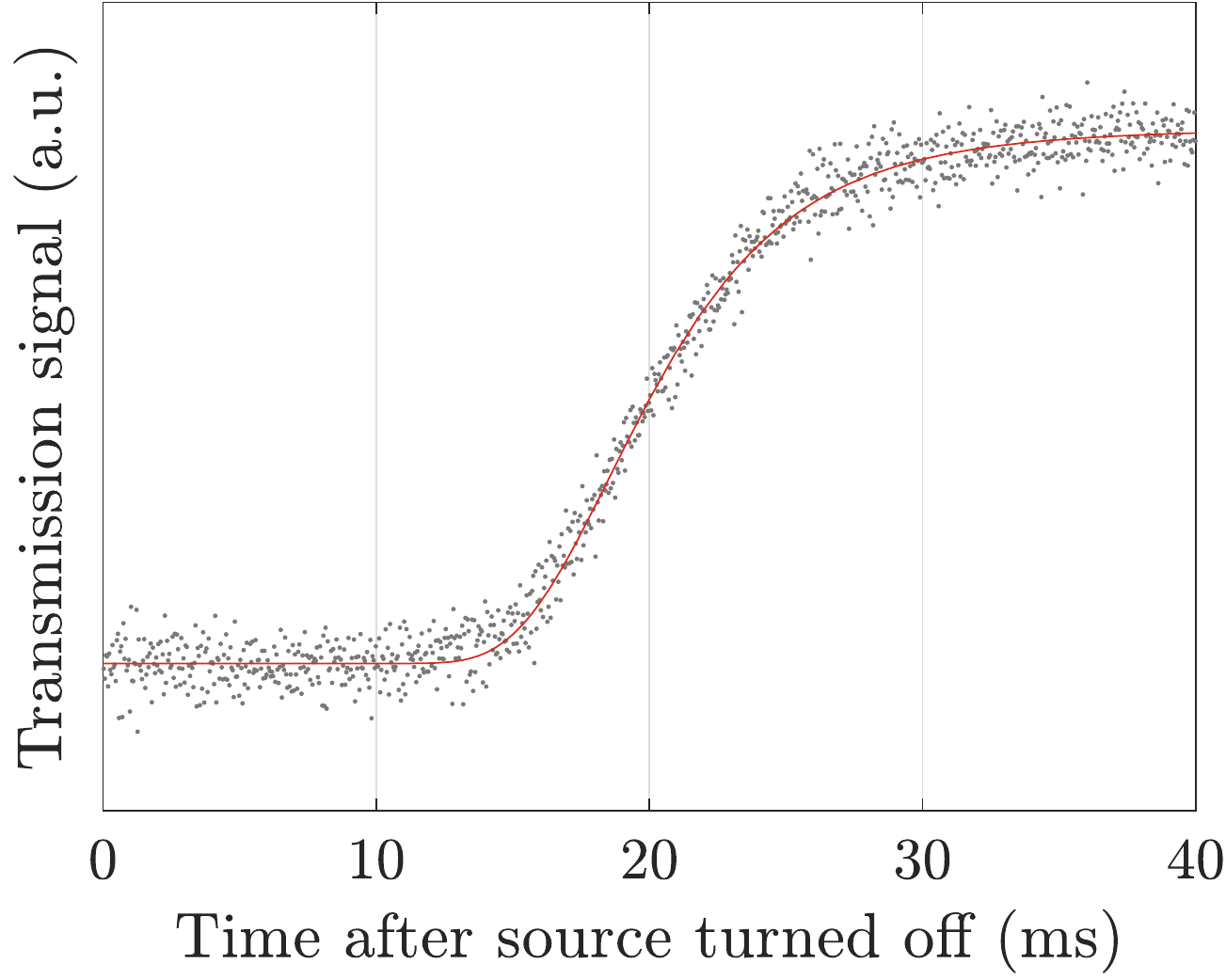}
					\quad\quad\quad
					\includegraphics[clip]{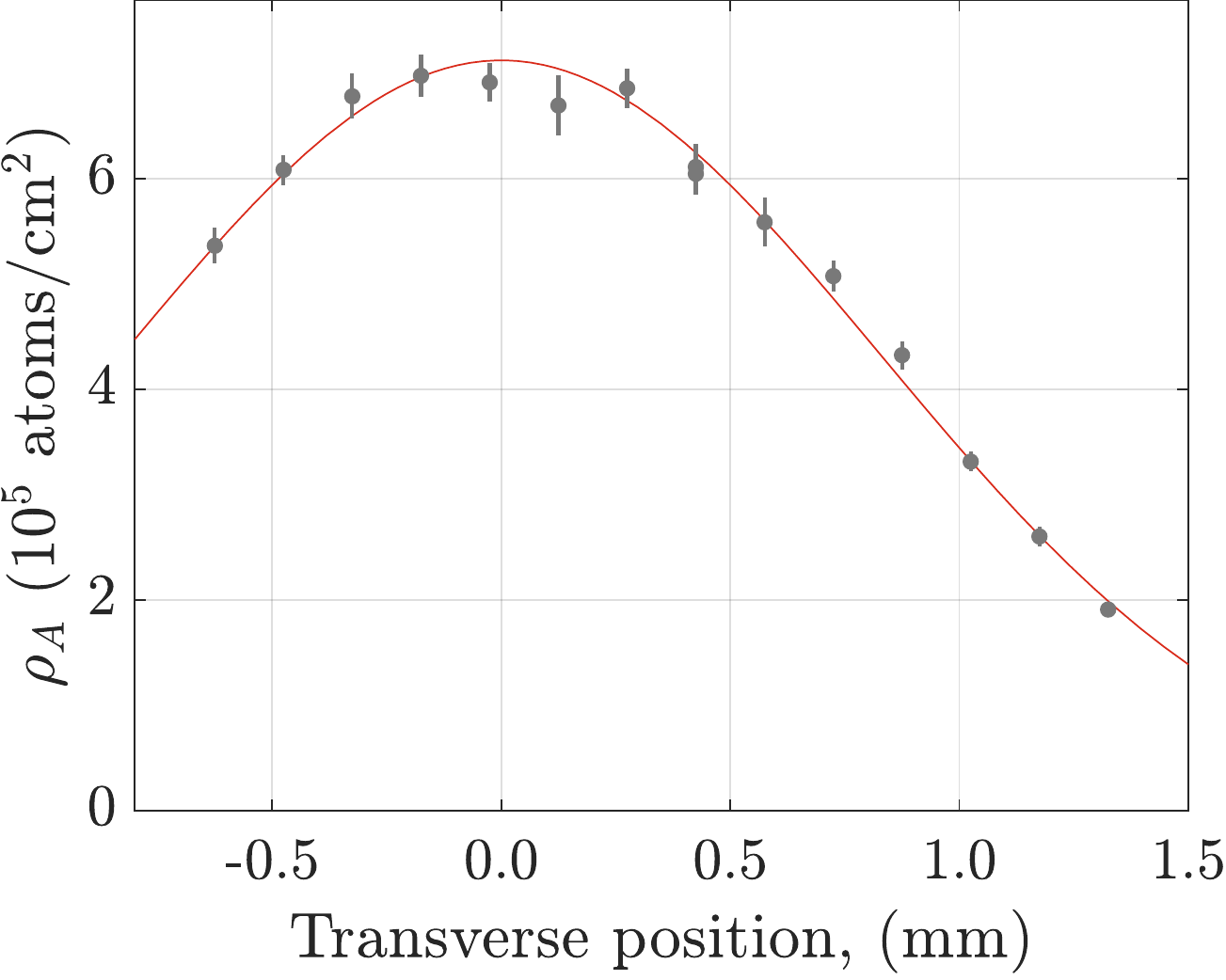}
				}
			};
			\end{scope}
			\node [anchor=west] (label1) at (0.8,4) {(a)};
			\node [anchor=west] (label2) at (10.9,4) {(b)};
		\end{tikzpicture}
	}
	\caption{Example data with fits: (a) transit time measurement data, with the fit to \eqr{e:TransitTimeFunction} that allows the atomic velocity distribution to be determined; (b) data for the atomic column density, $\rho_A$, with the fit to \eqr{e:FluxFunction}, from which the divergence is calculated.
	}
	\label{f:ExampleData}
\end{figure}

\subsection{Atomic Velocity Distribution}

The axial velocity distribution was measured using a transit time method: the atomic beam was abruptly turned off (switching time $\lesssim$10\,\textmu s) by switching off the EOM producing the repumping sidebands, while monitoring the absorption of the probe light through the atomic beam. Each velocity class of atoms in the beam takes a different length of time $t$ to propagate from the MOT region to the monitoring position (distance $D$ away). Thus the velocity distribution $f(v)$ of the atoms is mapped on to the absorption signal, $S(t)$, which has the functional form \cite{Park2012Cold, Rathod2013Continuous}
\begin{equation}\label{e:TransitTimeFunction}
	S(t) \propto \erf \left( \frac{\bar{v} t - D}{\sqrt{2} \, \sigma_v t} \right)
\end{equation}
where $\bar{v}$ and $\sigma_v$ are the mean and standard deviation of $f(v)$ respectively. This assumes $f(v)$ is a Gaussian distribution, which is justified since the velocities arise from stochastic photon scattering events. \lfir{f:ExampleData}a shows a typical transit time signal fitted with \eqr{e:TransitTimeFunction}, from which $\bar{v}$ and $\sigma_v$ are determined.

Note that a limitation of \eqr{e:TransitTimeFunction} is that it assumes a constant velocity for the atoms, whereas in reality atoms are accelerated from near-stationary over a short distance (after which the acceleration is negligible due to the atoms being significantly Doppler- and Zeeman-detuned from the light). This method of analysis therefore underestimates $\bar{v}$, and a more accurate approach would be to measure the transit time of the atoms to two (or more) different points far from the acceleration region, and assume a constant velocity between those points. However, such multi-point measurement did not have a sufficient signal-to-noise ratio for precise velocity determination, except at the highest atomic fluxes, so we used \eqr{e:TransitTimeFunction}, but comparison with the high-flux multi-point measurements indicates that \eqr{e:TransitTimeFunction} underestimates the mean velocity by 22\%.

\subsection{Atomic Density}

The atomic density profile was measured by translating the narrow probe beam transversely across the atomic beam and recording the variation in absorption (the 1.5\,mm $1/e^2$ diameter probe beam is $\sim$20 times narrower than the atomic beam). Let the atomic volumetric density at a point in the atomic beam be $\rho_V(x,y,z)$, where $z$ is the direction parallel to the beam, and the probe light is directed along $x$. From the measured intensity of the transmitted probe light, the Beer-Lambert law can be used to calculate the column density $\rho_A(y,z) = \int \rho_V(x,y,z)\,\dd x$, and a series of measurements for a range of $y$ produced data as shown in \fir{f:ExampleData}b. Integrating the fitted profile yields a value for $\rho_L(z) = \int \rho_V(y,z)\,\dd y$, which is the number of atoms per unit length along $z$. Assuming there is no loss of atoms as they propagate along $z$, then $\rho_L(z)$ is independent of $z$, hence we write $\rho_L(z) = \rho_L$.

The functional form of $\rho_A(y,z)$ to fit to the column density data (\fir{f:ExampleData}b) is Gaussian, since the column density profile arises from the stochastic photon scattering that generates the transverse velocity spread of the atoms. The explicit form of $\rho_A(y,z)$ depends on what fraction of the atomic beam width along $x$ is measurable. Ideally the probe light would pass through the entire width, however our measurements were conducted in a glass cell whose finite width cut off a fraction of the flux. For a measurement region of width $w$ along the measurement direction $x$ (in our case the inside dimension of the glass cell), the measured column density is $\rho_A(y,z) = \int_{-w/2}^{+w/2} \rho_V(x,y,z)\dd x$. Assuming the atomic beam profile has a radially symmetric Gaussian profile with standard deviation $\sigma_r$, this gives
\begin{equation}\label{e:FluxFunction}
	\rho_V(y,z) = \rho_L \erf \left( \frac{w}{2 \sqrt{2} \, \sigma_r} \right) \frac{e^{- (y - y_0)^2 / 2 \sigma_r^2}}{\sqrt{2 \pi} \, \sigma_r}
\end{equation}
Fitting this profile to the data provided an estimate of both $\sigma_r$, from which the atomic beam divergence was found, and also of $\rho_L$, which was combined with the atomic velocity distribution to calculate the total flux.

\subsection{Flux of Cold Atoms}\label{s:FluxEstimate}

The total flux was estimated from $\rho_L$ and the axial velocity distribution, $f(v)$. If $\Phi_0$ is the total flux (in \SI{}{atoms \per \second}), then the flux of atoms with velocity $v$ is $\Phi(v) = \Phi_0 f(v)$. The number of atoms per unit axial length with velocity $v$ is $\Phi(v) / v$, thus the total number of atoms per unit length, $\rho_L$, is
\begin{equation}
	\rho_L  = \int_{- \infty}^{\infty}\frac{\Phi(v)}{v} \, \dd v
			= \Phi_0 \int_{- \infty}^{\infty}\frac{f(v)}{v} \, \dd v
			= \Phi_0 \expval{\frac{1}{v}}
\end{equation}
and the total flux can therefore be calculated using $\Phi_0 = \rho_L / \expval{1 / v}$.

In general, $\expval{1 / v} \neq 1 / \bar{v}$, however, using the Taylor expansion of $\expval{1 / v}$ about the mean velocity $\bar{v}$,
\begin{equation}
	\expval{\frac{1}{v}} = \expval{ \frac{1}{\bar{v}} - \frac{v - \bar{v}}{v^2} + \frac{(v - \bar{v})^2}{v^3} - \cdots
	}
\end{equation}
then if $f(v)$ is a narrow distribution around a non-zero mean ($\bar{v} / \sigma_{v} \gg 1$), all higher order terms are small compared to the first, and $\expval{1 / v} \simeq 1 / \bar{v}$ is a good approximation. For our source $\bar{v} / \sigma_{v} \sim 3$, which causes $1 / \bar{v}$ to be an overestimate of $\expval{1 / v}$ by 22-23\%. This combined with the aforementioned underestimation arising from the velocity measurements leads to an overall systematic uncertainty in the flux which is estimated to be less than this (<22\%) when we estimate the total flux using $\Phi_0 \simeq \rho_L \bar{v}$.

\section{Results}\label{s:Results}

There were two aspects to characterising the atomic beam: i) determining the maximum flux to quantify the performance of our source and its applicability to practical quantum devices; and ii) exploring the effect of different parameters on the atomic beam, in order to better understand MOT sources. Specifically, we focussed on the impact of aperture size on the atomic beam, then looked at the effect of varying the intensity, laser beam diameter and MOT position (which is important for the robustness of portable devices to changes in the background magnetic field or mechanical vibrations).

\subsection{Maximum Flux}\label{s:MaxFlux}

The highest observed flux was $2.0(1) \times 10^{10}$\,atoms/s, for which the axial velocity distribution had a mean of 32(1)\,m/s and a FWHM of 27.6(9)\,m/s, and the FWHM divergence was 58(3)\,mrad (as recorded in Table~\ref{t:Sources}). This was obtained using the metal pyramid with a 1.38\,mm aperture, a $^{87}$Rb partial pressure of $1.3(1) \times 10^{-7}$\,mbar, and an intensity of 6.0\,$I_\textnormal{sat}$. This is the highest flux from a pyramid source reported in the literature to date, and we found that the intensity could be significantly reduced whilst still maintaining a high flux but producing a slower velocity. In particular, halving the intensity to 3.0\,$I_\textnormal{sat}$ reduced the flux by only 20\% to $1.6(1) \times 10^{10}$\,atoms/s with an average velocity of 22.6(7)\,m/s, velocity FWHM of 18.6(6)\,m/s and a divergence of 67(4)\,mrad.
The measured axial velocities were faster than most other sources, however we believe this to be straightforward to customise as required (see \secr{s:Intensity}).

Even operating at 50\% intensity, the atomic flux was still the highest reported from any pyramid, LVIS, 3D-MOT or grating source in the literature (see Table~\ref{t:Sources}). It is comparable or better than all but three known 2D or 2D$^{+}$ MOT sources (specifically \cite{Jollenbeck2011Hexapole, Park2012Cold, Schoser2002Intense}), and those all use glass cells, large cooling regions and four independent beams, which make them less suitable for portable applications. Note that although Schoser et al.\ \cite{Schoser2002Intense} obtained $6 \times 10^{10}$\,atoms/s \cite{Schoser2002Intense}, the velocity distribution (50\,m/s average, 75\,m/s FWHM) means only a fraction of the flux is capturable in a MOT. Also, Park et al.\ \cite{Park2012Cold}, reported that using two-colour pushing beams gave a twofold flux increase compared to a single push beam, making their flux directly comparable to ours, albeit in a less compact package.

\subsection{Flux vs Intensity}\label{s:Intensity}

\begin{figure}
	\centering
	\resizebox{0.9\textwidth}{!}{
		\includegraphics{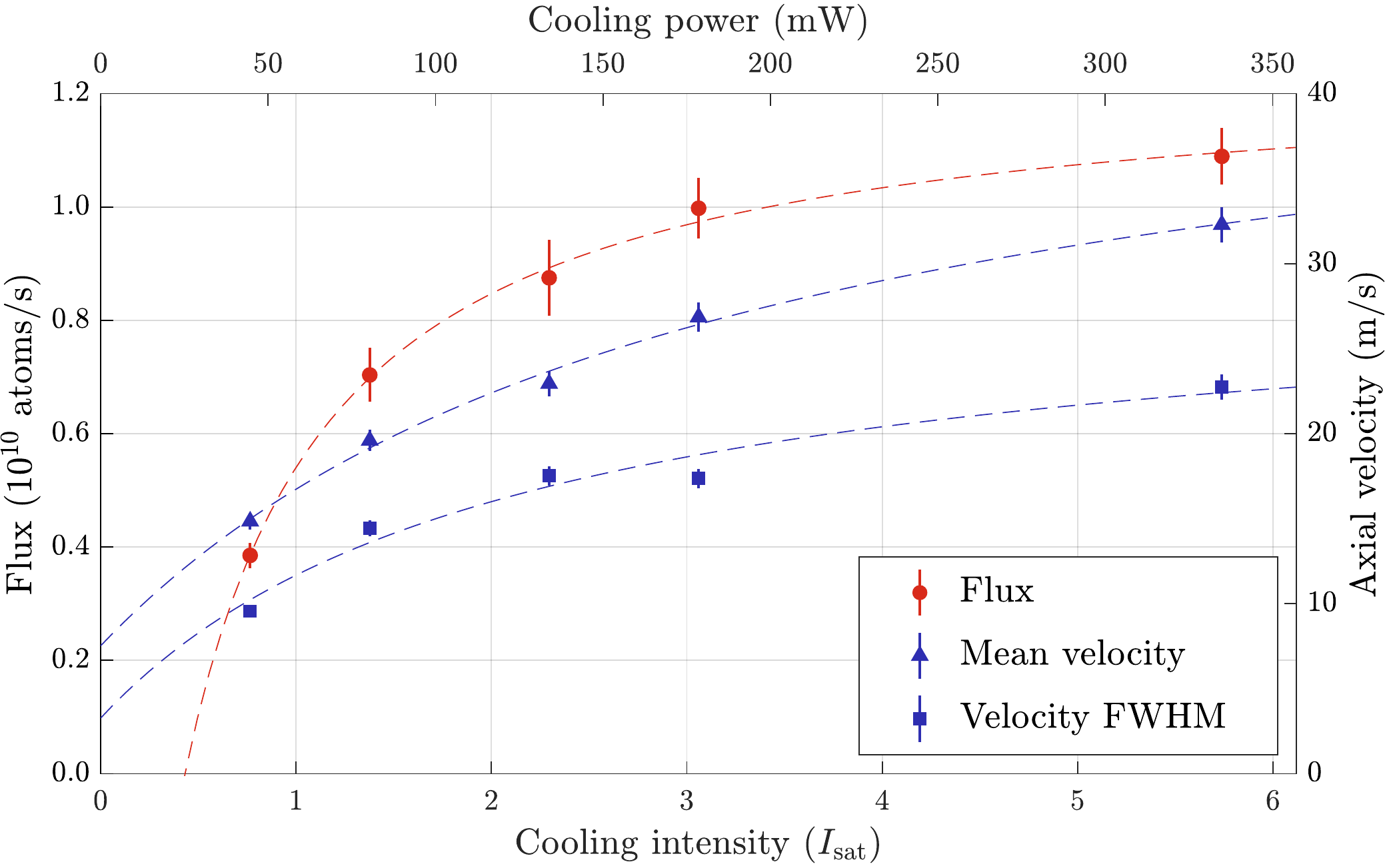}
	}
	\caption{The effect of the intensity of the single laser beam used for the pyramid MOT on the: flux (red circles), mean axial velocity (blue triangles), and velocity FWHM (blue, squares) of the atomic beam. This was obtained using the metal pyramid and a 1.90\,mm aperture. To guide the eye, fits to $a(I/b)/(1+I/b)+c$ are shown (based on the analytical form of the scattering rate \cite{Foot2005Atomic}).}
	\label{f:DataIntensity}
\end{figure}

The dependence of the flux and velocity of the atomic beam on the cooling light intensity is shown in \fir{f:DataIntensity}. The flux shows saturation with intensity, which allowed a high flux to be maintained despite significantly reducing the intensity. On the other hand, the average velocity and velocity spread do not saturate as quickly, with the result that altering the intensity has a strong effect on the velocity distribution. The velocity distribution depends on the intensity imbalance in the exit channel, so a pyramid source can be customised for a specific application by adjusting the intensity at the centre of the MOT beam. For example, compact devices often require slower atomic beams than laboratory systems, since the MOTs used to recapture the atoms are size-limited, with a correspondingly low capture velocity. By placing a mask in front of the pyramid that attenuates the centre of the MOT laser beam, the velocity of the atoms can be reduced (to 15\,m/s with 0.8\,$I_\textnormal{sat}$, for example, from \fir{f:DataIntensity}) without substantial change to the total flux, provided the velocity is not slow enough for a significant proportion of atoms to be lost to background collisions before reaching the recapture MOT. In addition, Lu et al.\ found that altering the MOT beam polarisation allowed the velocity distribution to be modified (particularly the FWHM spread) without a substantial flux reduction \cite{Lu1996Low}, which opens another approach to customisation. Thus the pyramid source is straightforwardly adjustable whilst remaining compact for practical cold-atom devices.

\subsection{Flux vs Aperture Size}\label{s:Aperture}

We investigated the effect of aperture size on the atomic beam, and the results are shown in \fir{f:DataAperture} for both the glass and metal pyramids. For each aperture size, the background $^{87}$Rb pressure was varied (see the inset of \fir{f:DataAperture}a) and the peak flux was recorded (the optimum pressure was typically 1\,-\,$3 \times 10^{-7}$\,mbar). Initial measurements were taken using the glass pyramid, however the absolute flux was limited due to abnormally high scatter from the dielectric mirror coatings, so measurements were repeated with the metal pyramid to better estimate the maximum flux. The metal pyramid produced higher fluxes, as well as significantly higher axial velocities (shown in \fir{f:DataAperture}b) because the scattered light in the glass pyramid reduced the axial intensity imbalance that accelerated the atoms. Reducing the cooling light intensity for the metal pyramid by 50\% still produced larger fluxes than the glass pyramid, but with similar atomic velocities.

The data in \fir{f:DataAperture}a shows a clear increase in flux with aperture size up to about \SI{1.2}{\milli\metre} before flattening off. For both dielectric and metal mirrors we found the flux with \SI{2}{\milli\metre} aperture was lower than the peak flux. We hypothesise that this behaviour is due to the relative magnitude of the size of the exit channel and the transverse velocities that atoms enter the exit channel with. For very large apertures the exit channel is wide, extending significantly away from the axis of the pyramid, meaning that atoms are not fully cooled before entering the exit channel, and hence have significant transverse velocities which allow them to traverse the channel before reaching the aperture. They are then lost from the atomic beam, causing the reduction in flux for large apertures. At the other extreme, very small apertures form a narrow channel which can be traversed by atoms with even very low transverse velocities, resulting in a reduced flux for these very small apertures. Note that a consequence of a small aperture (and hence a narrow channel) is that atoms take longer to enter the channel, which we were able to observe by the presence of significant fluorescence; for apertures below 1\,mm, a cloud of fluorescing atoms was clearly visible in the pyramid (similar to that in a standard MOT), indicating that atoms entered the exit channel at a slower rate than the MOT capture rate (which is the theoretical maximum flux).

The flux appears to saturate with aperture size (instead of a strongly peaked dependence) which is, we hypothesise, due to recycling of atoms \cite{Lu1996Low, Wohlleben2001Atom, Kohel2003Generation}; atoms that escape from the exit channel can be recaptured and directed back to the exit channel. An atom can be recycled multiple times, but will eventually be lost due to background collisions. Without recycling, the dependence of the flux on the aperture size would be expected to be more strongly peaked.

\begin{figure}
	\centering
	\resizebox{0.9\textwidth}{!}{
		\begin{tikzpicture}
		\begin{scope}
		\node[anchor=south west,inner sep=0] (image) at (0,0) {
			\resizebox{0.9\textwidth}{!}{
				\includegraphics{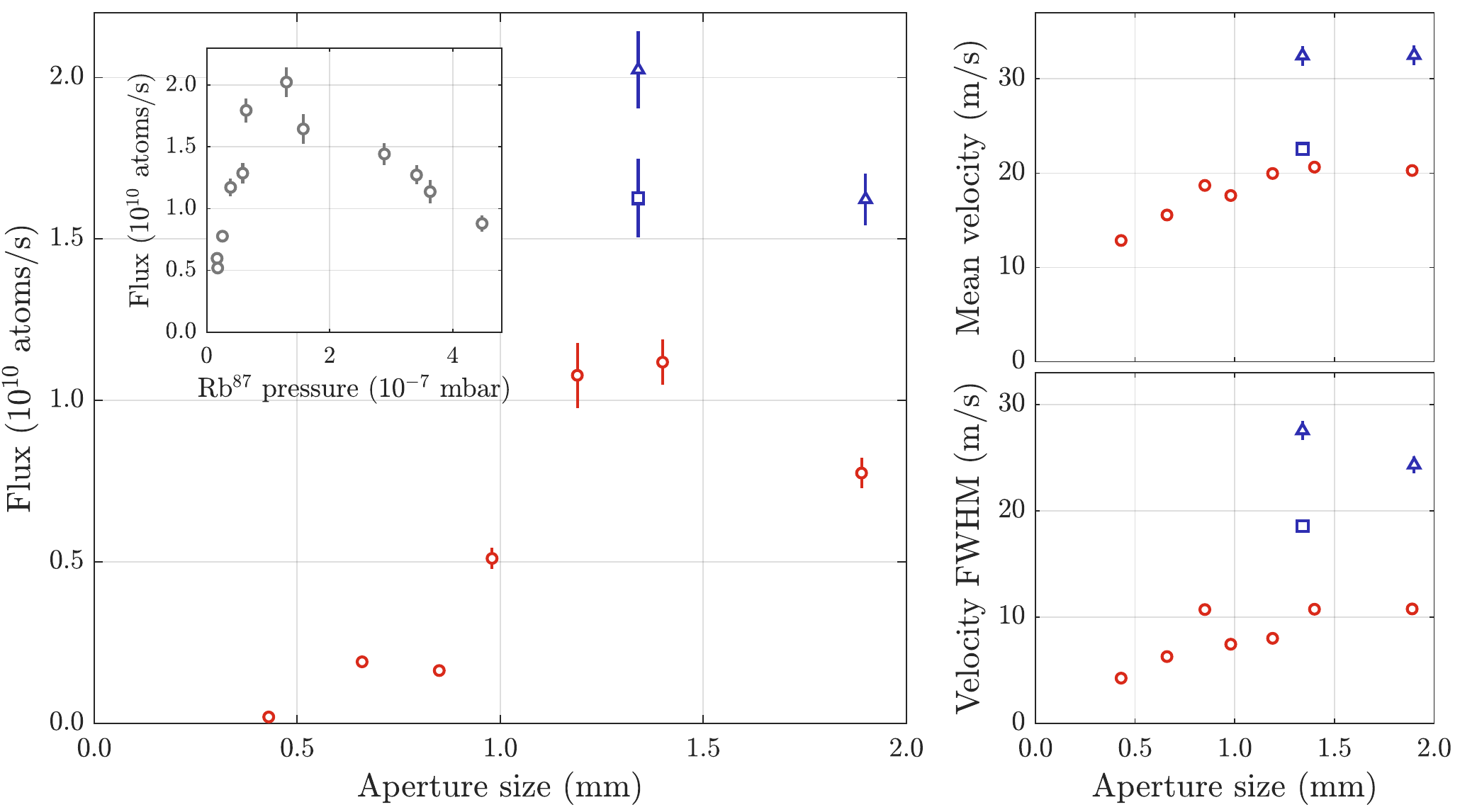}
			}
		};
		\end{scope}
		\node [anchor=west] (label1) at (6.67, 6.05) {(a)};
		\node [anchor=west] (label2) at (8.50, 6.05) {(b)};
		\node [anchor=west] (label3) at (8.50, 3.12) {(c)};
		\end{tikzpicture}
	}
	\caption{The effect of aperture size on the: (a) flux, (b) mean axial velocity and (c) velocity FWHM of the atomic beam. Data taken with the glass pyramid (red circles) did not reveal the maximum possible flux due to high scatter from the dielectric mirror coating, so data was taken with the metal pyramid (blue triangles) to better establish the maximum flux. Data is also shown for the metal pyramid operating with 50\% intensity (blue squares), for which the flux is only reduced by 26\%. The inset of (a) shows the effect on the flux of varying pressure on for a constant aperture size (1.38\,mm in this case).}
	\label{f:DataAperture}
\end{figure}
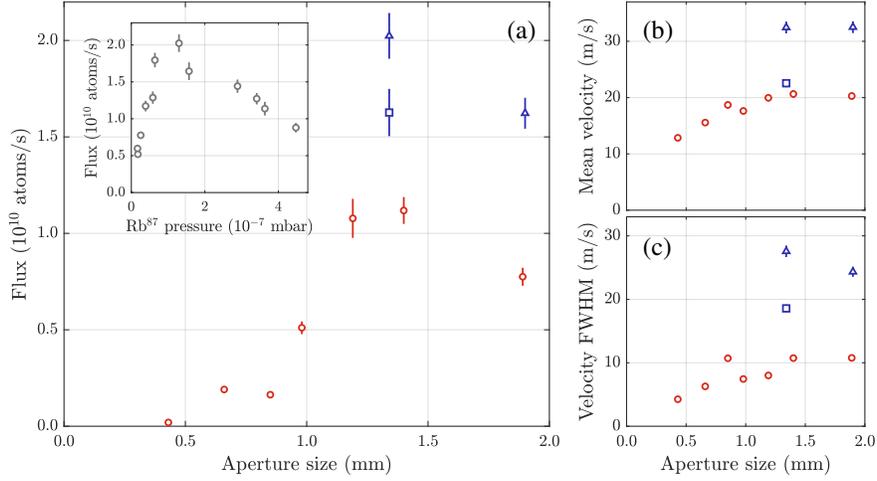

The mean velocity and velocity FWHM showed no clear trend with aperture size, other than a small decrease of velocity for small apertures (\fir{f:DataAperture}b). This is likely due to a combination of two effects. Firstly, atoms are cooled further before reaching the exit channel for a smaller aperture (as the aperture size tends to zero the system tends towards a MOT with a small leak), so enter the exit channel with a lower velocity. Secondly, there is diffraction of the retroreflected light by the aperture (in the same way a collimated beam diffracts around an opaque mask), which results in some retroreflected light being angled towards the pyramid axis, rather than being parallel to it. The exit channel is therefore not indefinitely long, but instead tapers to a point where the diffracted light fills the channel. For large apertures this diffraction is negligible, but for small apertures it is sufficient to reduce the intensity imbalance that the atoms experience during acceleration from the MOT centre, reducing the velocity.

\subsection{Pyramid Diameter}

To determine whether the flux was limited by the pyramid size, the MOT laser beam diameter was varied using an adjustable iris to simulate the performance of smaller diameter pyramids (data not included here). We observed an approximately linear increase with diameter for multiple aperture sizes. This demonstrates that a larger pyramid would be expected to produce an even higher flux, and we are currently exploring this avenue for increasing the flux.

Note that the flux from a MOT source is in theory expected to vary with the beam diameter $d$ as $D^{3.6}$ \cite{Gibble1992Improved, Lindquist1992Experimental, Hoth2013Atom}, but that assumes a variable Gaussian beam diameter, rather than apodising a fixed-diameter Gaussian beam with an iris, as in this case.

\subsection{MOT Position \& Aperture Size}

Transverse and axial magnetic bias fields were applied to determine the effect on the atomic beam of changing the position of the MOT field centre. In both cases the divergence and axial velocity distribution were found to be largely unchanged. This is because these properties only depend on the light field that accelerates the atoms.

The effect on the flux is shown in \fir{f:DataBiasFields}. The flux was significantly more sensitive to transverse fields than axial fields; a 20\% decrease in flux required a 1\,G axial field, but just a 0.2\,-\,0.3\,G transverse field. This is due to a combination of two factors. Firstly, the transverse gradient is half of the axial gradient for a 3D quadrupole field, so a given bias field causes a larger movement of the MOT when applied transversely. Secondly, the atomic beam is produced by transferring trapped atoms into the exit channel along the pyramid axis, and while an axial bias field just moves the MOT approximately (depending on the uniformity of the background field) parallel to the exit channel, a transverse bias field moves the MOT perpendicularly away from the channel, causing a more substantial reduction of how efficiently trapped atoms are transferred to the exit channel, and therefore a reduced flux.

In addition to the difference between axial and transverse fields, different aperture sizes affected the sensitivity of the flux to external fields. For a small (\SI{0.66}{\milli\metre}) aperture, the dependence on transverse field was sharply peaked \fir{f:DataBiasFields}a, whereas for a large (\SI{1.89}{\milli\metre}) aperture, the transverse field dependence plateaued, because for a larger aperture, small movements of the MOT have a reduced impact on the overlap of the trapped atoms with the exit channel. Similarly, \fir{f:DataBiasFields}b shows the large aperture was also more robust to axial bias fields. This makes a larger aperture favourable in portable applications where the MOT position may vary by more than \SI{100}{\micro\metre}, either due to changing magnitude and direction of the ambient magnetic field (e.g. the Earth's background magnetic field is 0.25-0.65\,G), or due to mechanical vibrations or misalignment of the magnetic-field coils relative to the pyramid.

\begin{figure}
	\centering
	\resizebox{0.9\textwidth}{!}{
		\begin{tikzpicture}
			\begin{scope}
			\node[anchor=south west,inner sep=0] (image) at (0,0) {
				\resizebox{0.9\textwidth}{!}{
					\includegraphics{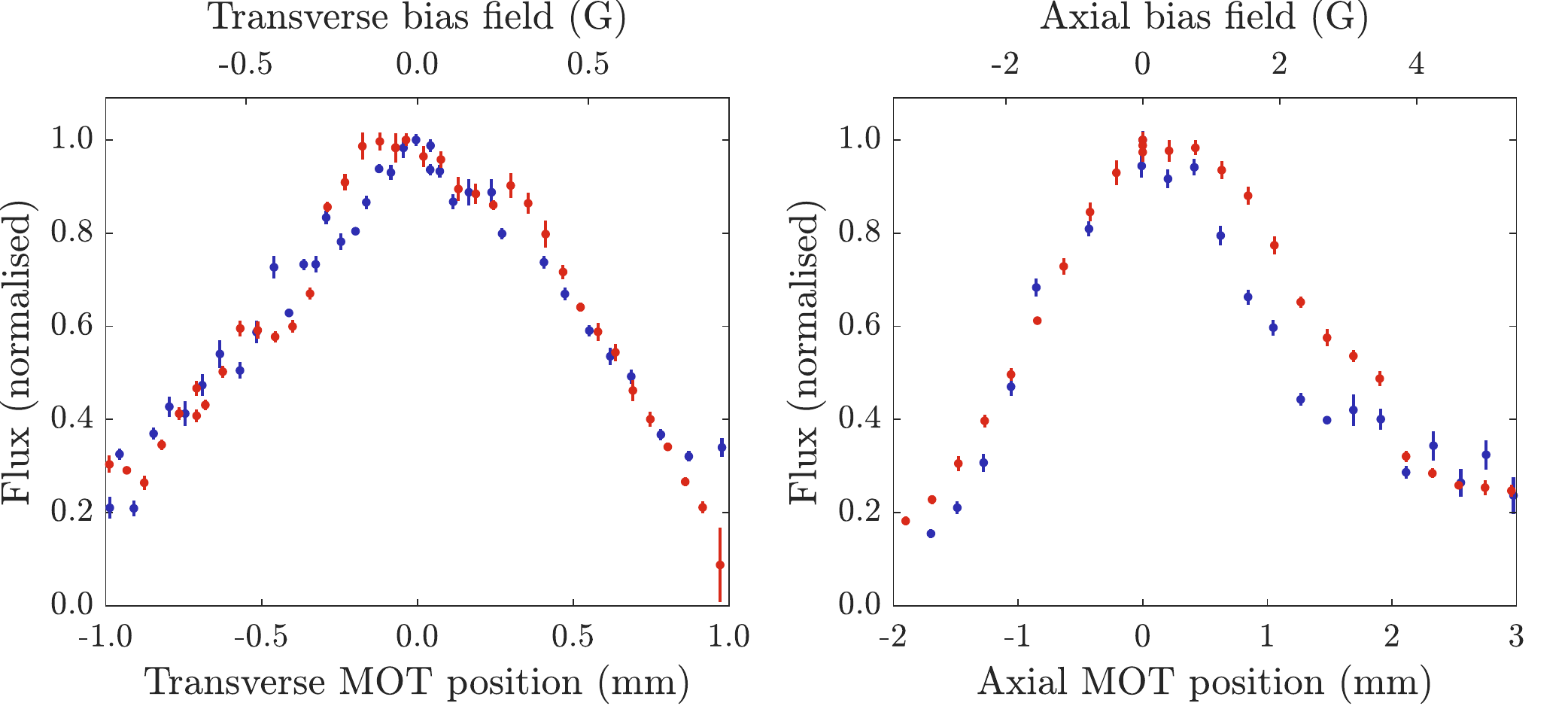}
				}
			};
			\end{scope}
			\node [anchor=west] (label1) at (1.0,4.2) {(a)};
			\node [anchor=west] (label2) at (6.98,4.2) {(b)};
		\end{tikzpicture}
	}
	\caption[Flux dependence on the MOT position \& aperture size]{The normalised flux from the glass pyramid as the position of the MOT field centre was changed, caused by applying transverse (a) and axial (b) bias fields for aperture sizes of \SI{0.66}{\milli\metre} (blue) and \SI{1.89}{\milli\metre} (red). The sensitivity of the MOT position to the applied transverse/axial bias field was \SI{1.10/0.55}{\milli \metre \per G}, and negative axial bias fields moved the MOT position towards the pyramid aperture. Applied fields are shown relative to the optimal values (9.23\,G axially, and 0.98\,G/0.51\,G transversely for the \SI{0.66}{\milli \metre}/\SI{1.89}{\milli \metre} apertures respectively).}
	\label{f:DataBiasFields}
\end{figure}

\section{Summary \& Outlook}

We have developed and tested a compact, high-flux cold-atom source suitable for portable quantum technology devices using cold atoms. The measured flux of $2.0(1) \times 10^{10}$\,atoms/s was greater or comparable to all other MOT sources in the literature, making our design an excellent candidate for not just portable devices, but also research laboratory systems. Even halving the optical power to 195\,mW only reduced the flux by 26\% to $1.6(1) \times 10^{10}$\,atoms/s. The design is scalable and a larger pyramid is expected to produce an even higher flux. Conversely, the capture rate is sufficiently high that the design if scaled down would still produce useful fluxes for applications where size is the priority rather than high fluxes. The velocity of the atomic beam was higher than many other sources but, if necessary, could be easily tuned by attenuating the centre of the MOT beam. An alternative would be to add a plug beam to reduce the intensity imbalance in the exit channel, or to use two-colour pushing beams, which would give far greater control over the velocity distribution and a potential two-fold increase in flux \cite{Park2012Cold}, although would significantly alter the single-beam simplicity that makes the pyramid design so suitable for compact applications.

The unique adjustable aperture feature of our source enabled us to observe a strong dependence of the flux on aperture size, with an optimal aperture size of about 1.4$\pm$0.2\,mm. We also found that larger aperture sizes made the source more robust to changes in the MOT position. Larger aperture sizes therefore appear better suited to portable devices subject to mechanical stress or changing magnetic fields, such as repositioning or reorientation in the geomagnetic field.

The adjustable aperture would provide additional benefits if developed to be adjusted in-vacuum; the aperture could function as a mechanical shutter, or enable a variable differential pumping rate (perhaps to speed up vacuum bake-outs). The all-metal pyramid design also enables potentially higher bake-out temperatures than designs using glass mirrors and adhesive, with the sole limitation being the temperature stability of the dielectric mirror coating.

To further improve the flux, we are investigating a Zeeman-pyramid hybrid source, where the magnetic field profile is designed such that atoms outside the pyramid experience a Zeeman-slower-type field (as illustrated in \fir{f:ZeemanHybrid}), which should increase the capture velocity of the pyramid MOT. Even a small increase in capture velocity would result in a significant increase in capture rate, and hence in flux, due to the Maxwell-Boltzmann distribution of atomic velocities. Another as yet unexplored parameter is the angle of the pyramid mirrors; if the mirrors were angled to direct the retroreflected light slightly towards the axis, the exit channel would be narrowed at the position of the MOT (as shown in \fir{f:ZeemanHybrid}(a)). Atoms would then always enter the exit region with very low transverse velocities, even for large apertures, enabling access to even higher fluxes with minimal changes to our compact design.

\begin{figure}
	\centering
	\resizebox{0.9\textwidth}{!}{
		\begin{tikzpicture}
		\begin{scope}
		\node[anchor=south west,inner sep=0] (image) at (0,0) {
			\resizebox{0.9\textwidth}{!}{
				\includegraphics{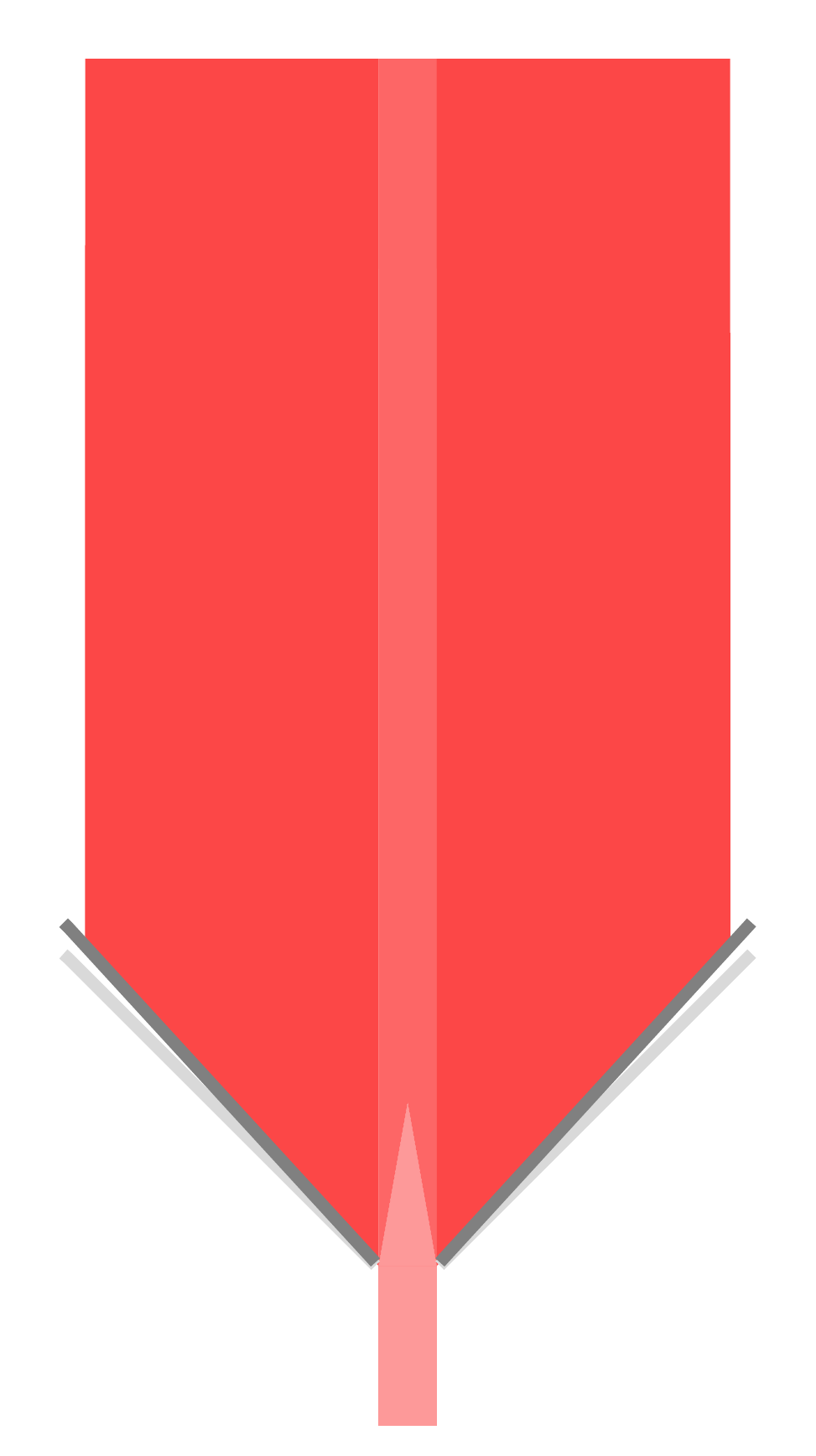}
				\quad\quad\quad
				\includegraphics{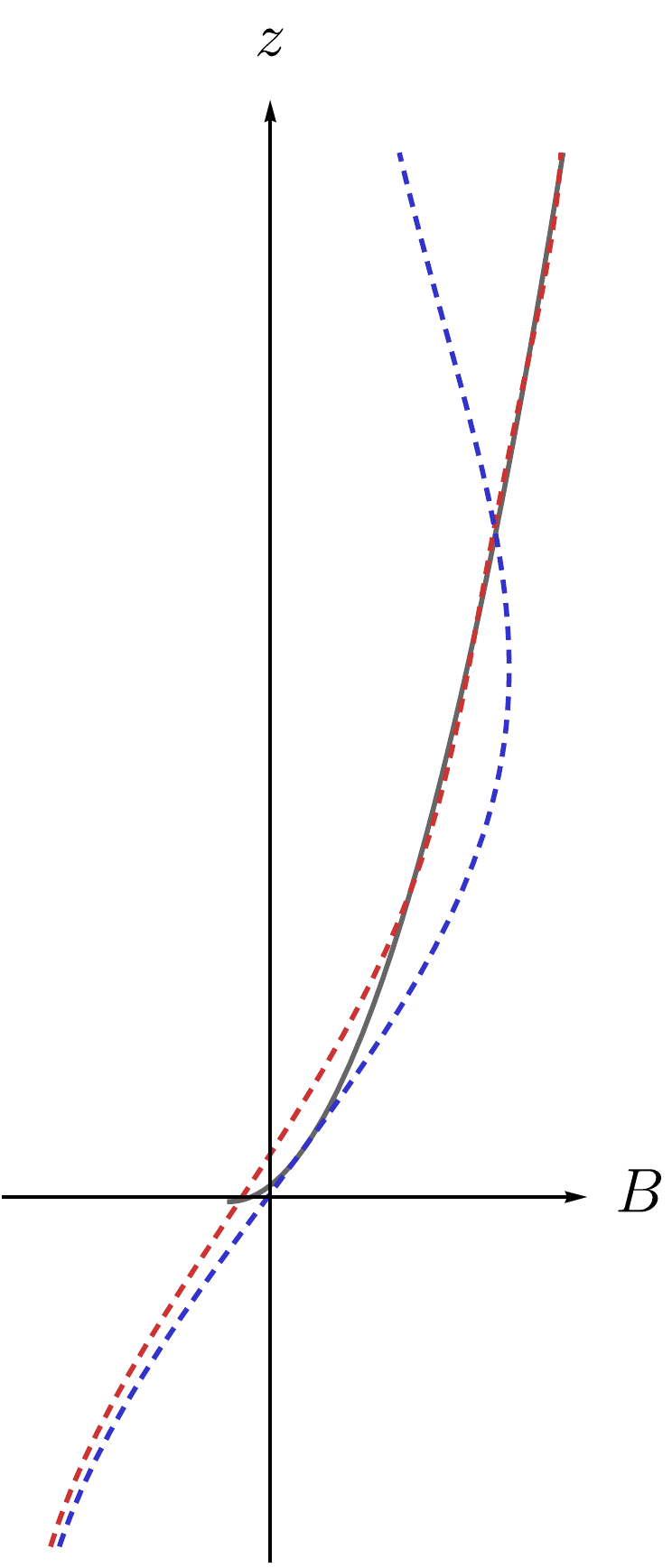}
				\quad\quad\quad\quad
				\includegraphics{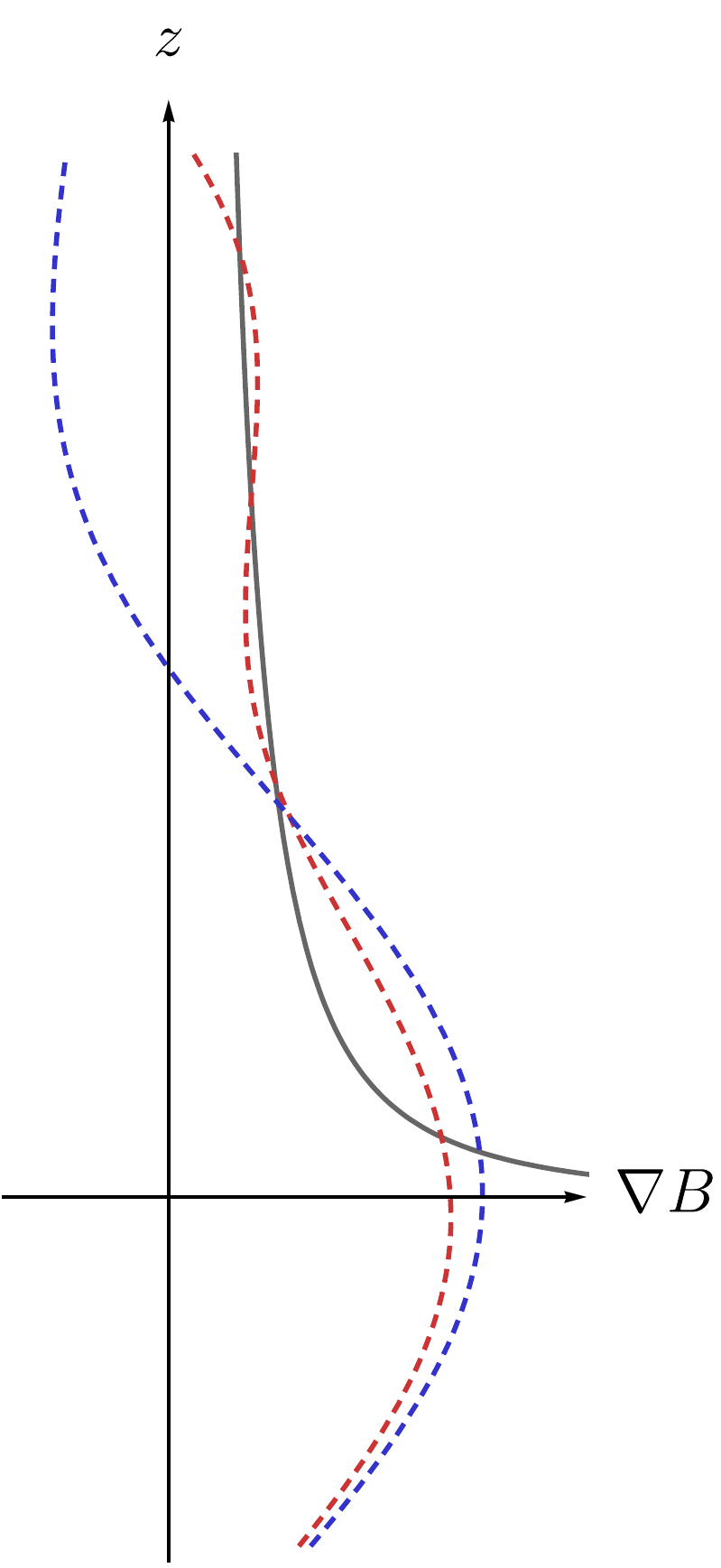}
			}
		};
		\end{scope}
		\node [anchor=west,white] (label1) at (0.7,6.6) {(a)};
		\node [anchor=west] (label2) at (4.8,6.6) {(b)};
		\node [anchor=west] (label3) at (10.6,6.6) {(c)};
		\end{tikzpicture}
	}
	\caption{(a): A pyramid with mirrors at an acuter angle than standard (90\textdegree\ between opposite mirrors, shown as pale grey), which causes the exit channel (formed by the absence of retroreflected light) to taper to a point, instead of extending along the axis. This would prevent atoms entering the exit channel before being fully cooled, and increase the flux by reducing the number of atoms that escape the exit channel before reaching the aperture. (b) \& (c): The magnetic field and magnetic field gradient profiles for a MOT (blue, dashed), a Zeeman slower (grey, solid), and a proposed MOT-Zeeman hybrid (red,dashed) that extends the cooling region out of the pyramid along the axis. Note that (a) shares the same $z$-axis as (b) and (c).}
	\label{f:ZeemanHybrid}
\end{figure}

\section*{Funding}

We gratefully acknowledge funding and support from the Defence Science and Technology Laboratory (Dstl), UK, and from the following grants administered by the Engineering and Physical Sciences Research Council (EPSRC), UK: the Networked Quantum Information Technologies hub (EP/T001062); the UK Quantum Technology Hub for Sensors and Metrology (EP/M013294); Innovate UK (EP/R001685/1, jointly with the company M Squared Lasers, UK); the EPSRC Impact Acceleration Accounts (IAA) Technology Fund (EP/R511742); and the EPSRC IAA Doctoral Impact Scheme.

\section*{Acknowledgements}

The authors appreciate fruitful discussions with Elliot Bentine.

\section*{Disclosures}

The authors declare no conflicts of interest.

\section*{Supplementary Material}

CAD files in various formats for the glass and metal pyramids presented in this paper are available at the Oxford University Research Archive, accessible at \href{https://ora.ox.ac.uk/objects/uuid:2ac612c7-7045-4757-a490-05e61ca551b2}{ora.ox.ac.uk/objects/uuid:2ac612c7-7045-4757-a490-05e61ca551b2}, and more details are available upon request from the authors.

\bibliography{References}

\begin{thebibliography}{10}
\newcommand{\enquote}[1]{``#1''}

\bibitem{Schioppo2017Ultrastable}
M.~Schioppo, R.~C. Brown, W.~F. McGrew, N.~Hinkley, R.~J. Fasano, K.~Beloy,
  T.~H. Yoon, G.~Milani, D.~Nicolodi, J.~A. Sherman, N.~B. Phillips, C.~W.
  Oates, and A.~D. Ludlow, \enquote{{Ultrastable optical clock with two
  cold-atom ensembles},} {\protect\JournalTitle{Nature Photonics}} \textbf{11},
  48--52 (2017).

\bibitem{Poli2014Transportable}
N.~Poli, M.~Schioppo, S.~Vogt, S.~Falke, U.~Sterr, C.~Lisdat, and G.~M. Tino,
  \enquote{{A transportable strontium optical lattice clock},}
  {\protect\JournalTitle{Applied Physics B}} \textbf{117}, 1107--1116 (2014).

\bibitem{Ludlow2015Optical}
A.~D. Ludlow, M.~M. Boyd, J.~Ye, E.~Peik, and P.~O. Schmidt, \enquote{{Optical
  atomic clocks},} {\protect\JournalTitle{Reviews of Modern Physics}}
  \textbf{87}, 637--701 (2015).

\bibitem{Elvin2019Cold}
R.~Elvin, G.~W. Hoth, M.~Wright, B.~Lewis, J.~P. McGilligan, A.~S. Arnold,
  P.~F. Griffin, and E.~Riis, \enquote{{Cold-atom clock based on a diffractive
  optic},} {\protect\JournalTitle{Optics Express}} \textbf{27}, 38359 (2019).

\bibitem{Bowden2019Pyramid}
W.~Bowden, R.~Hobson, I.~R. Hill, A.~Vianello, M.~Schioppo, A.~Silva, H.~S.
  Margolis, P.~E. Baird, and P.~Gill, \enquote{{A pyramid MOT with integrated
  optical cavities as a cold atom platform for an optical lattice clock},}
  {\protect\JournalTitle{Scientific Reports}} \textbf{9}, 11704 (2019).

\bibitem{Grotti2018Geodesy}
J.~Grotti, S.~Koller, S.~Vogt, S.~H{\"{a}}fner, U.~Sterr, C.~Lisdat, H.~Denker,
  C.~Voigt, L.~Timmen, A.~Rolland, F.~N. Baynes, H.~S. Margolis, M.~Zampaolo,
  P.~Thoumany, M.~Pizzocaro, B.~Rauf, F.~Bregolin, A.~Tampellini, P.~Barbieri,
  M.~Zucco, G.~A. Costanzo, C.~Clivati, F.~Levi, and D.~Calonico,
  \enquote{{Geodesy and metrology with a transportable optical clock},}
  {\protect\JournalTitle{Nature Physics}} \textbf{14}, 437--441 (2018).

\bibitem{Bodart2010Cold}
Q.~Bodart, S.~Merlet, N.~Malossi, F.~P. {Dos Santos}, P.~Bouyer, and
  A.~Landragin, \enquote{{A cold atom pyramidal gravimeter with a single laser
  beam},} {\protect\JournalTitle{Applied Physics Letters}} \textbf{96}, 134101
  (2010).

\bibitem{Bidel2013Compact}
Y.~Bidel, O.~Carraz, R.~Charri{\`{e}}re, M.~Cadoret, N.~Zahzam, and A.~Bresson,
  \enquote{{Compact cold atom gravimeter for field applications},}
  {\protect\JournalTitle{Applied Physics Letters}} \textbf{102}, 144107 (2013).

\bibitem{Gillot2016LNE}
P.~Gillot, B.~Cheng, A.~Imanaliev, S.~Merlet, and F.~{Pereira Dos Santos},
  \enquote{{The LNE-SYRTE cold atom gravimeter},} in \emph{2016 European
  Frequency and Time Forum (EFTF),}  (IEEE, 2016), pp. 1--3.

\bibitem{Menoret2018Gravity}
V.~M{\'{e}}noret, P.~Vermeulen, N.~{Le Moigne}, S.~Bonvalot, P.~Bouyer,
  A.~Landragin, and B.~Desruelle, \enquote{{Gravity measurements below 10-9 g
  with a transportable absolute quantum gravimeter},}
  {\protect\JournalTitle{Scientific Reports}} \textbf{8}, 12300 (2018).

\bibitem{Bidel2018Absolute}
Y.~Bidel, N.~Zahzam, C.~Blanchard, A.~Bonnin, M.~Cadoret, A.~Bresson,
  D.~Rouxel, and M.~Lequentrec-Lalancette, \enquote{{Absolute marine gravimetry
  with matter-wave interferometry},} {\protect\JournalTitle{Nature
  Communications}} \textbf{9}, 627 (2018).

\bibitem{Freier2016Mobile}
C.~Freier, M.~Hauth, V.~Schkolnik, B.~Leykauf, M.~Schilling, H.~Wziontek, H.-G.
  Scherneck, J.~M{\"{u}}ller, and A.~Peters, \enquote{{Mobile quantum gravity
  sensor with unprecedented stability},} {\protect\JournalTitle{Journal of
  Physics: Conference Series}} \textbf{723}, 012050 (2016).

\bibitem{Wu2019Gravity}
X.~Wu, Z.~Pagel, B.~S. Malek, T.~H. Nguyen, F.~Zi, D.~S. Scheirer, and
  H.~M{\"{u}}ller, \enquote{{Gravity surveys using a mobile atom
  interferometer},} {\protect\JournalTitle{Science Advances}} \textbf{5},
  eaax0800 (2019).

\bibitem{Wang2018Shift}
S.-K. Wang, Y.~Zhao, W.~Zhuang, T.-C. Li, S.-Q. Wu, J.-Y. Feng, and C.-J. Li,
  \enquote{{Shift evaluation of the atomic gravimeter NIM-AGRb-1 and its
  comparison with FG5X},} {\protect\JournalTitle{Metrologia}} \textbf{55},
  360--365 (2018).

\bibitem{Mazon2019Portable}
M.~J. Mazon, G.~H. Iyanu, and H.~Wang, \enquote{{A portable, compact cold atom
  physics package for atom interferometry},} in \emph{2019 Joint Conference of
  the IEEE International Frequency Control Symposium and European Frequency and
  Time Forum (EFTF/IFC),}  (IEEE, 2019), pp. 1--5.

\bibitem{Geiger2011Detecting}
R.~Geiger, V.~M{\'{e}}noret, G.~Stern, N.~Zahzam, P.~Cheinet, B.~Battelier,
  A.~Villing, F.~Moron, M.~Lours, Y.~Bidel, A.~Bresson, A.~Landragin, and
  P.~Bouyer, \enquote{{Detecting inertial effects with airborne matter-wave
  interferometry},} {\protect\JournalTitle{Nature Communications}} \textbf{2},
  474 (2011).

\bibitem{Wu2017Multiaxis}
X.~Wu, F.~Zi, J.~Dudley, R.~J. Bilotta, P.~Canoza, and H.~M{\"{u}}ller,
  \enquote{{Multiaxis atom interferometry with a single-diode laser and a
  pyramidal magneto-optical trap},} {\protect\JournalTitle{Optica}} \textbf{4},
  1545 (2017).

\bibitem{Battelier2016Development}
B.~Battelier, B.~Barrett, L.~Fouch{\'{e}}, L.~Chichet, L.~Antoni-Micollier,
  H.~Porte, F.~Napolitano, J.~Lautier, A.~Landragin, and P.~Bouyer,
  \enquote{{Development of compact cold-atom sensors for inertial navigation},}
  in \emph{Quantum Optics,}  J.~Stuhler and A.~J. Shields, eds. (2016), p.
  990004.

\bibitem{Cheiney2018Navigation}
P.~Cheiney, L.~Fouch{\'{e}}, S.~Templier, F.~Napolitano, B.~Battelier,
  P.~Bouyer, and B.~Barrett, \enquote{{Navigation-compatible hybrid quantum
  accelerometer using a Kalman filter},} {\protect\JournalTitle{Physical Review
  Applied}} \textbf{10}, 034030 (2018).

\bibitem{Becker2018Space}
D.~Becker, M.~D. Lachmann, S.~T. Seidel, H.~Ahlers, A.~N. Dinkelaker,
  J.~Grosse, O.~Hellmig, H.~M{\"{u}}ntinga, V.~Schkolnik, T.~Wendrich,
  A.~Wenzlawski, B.~Weps, R.~Corgier, T.~Franz, N.~Gaaloul, W.~Herr,
  D.~L{\"{u}}dtke, M.~Popp, S.~Amri, H.~Duncker, M.~Erbe, A.~Kohfeldt,
  A.~Kubelka-Lange, C.~Braxmaier, E.~Charron, W.~Ertmer, M.~Krutzik,
  C.~L{\"{a}}mmerzahl, A.~Peters, W.~P. Schleich, K.~Sengstock, R.~Walser,
  A.~Wicht, P.~Windpassinger, and E.~M. Rasel, \enquote{{Space-borne
  Bose–Einstein condensation for precision interferometry},}
  {\protect\JournalTitle{Nature}} \textbf{562}, 391--395 (2018).

\bibitem{Elliott2018NASA}
E.~R. Elliott, M.~C. Krutzik, J.~R. Williams, R.~J. Thompson, and D.~C.
  Aveline, \enquote{{NASA's Cold Atom Lab (CAL): system development and ground
  test status},} {\protect\JournalTitle{npj Microgravity}} \textbf{4}, 16
  (2018).

\bibitem{Trimeche2019Concept}
A.~Trimeche, B.~Battelier, D.~Becker, A.~Bertoldi, P.~Bouyer, C.~Braxmaier,
  E.~Charron, R.~Corgier, M.~Cornelius, K.~Douch, N.~Gaaloul, S.~Herrmann,
  J.~M{\"{u}}ller, E.~Rasel, C.~Schubert, H.~Wu, and F.~{Pereira dos Santos},
  \enquote{{Concept study and preliminary design of a cold atom interferometer
  for space gravity gradiometry},} {\protect\JournalTitle{Classical and Quantum
  Gravity}} \textbf{36}, 215004 (2019).

\bibitem{Chiow2015Laser}
S.-w. Chiow, J.~Williams, and N.~Yu, \enquote{{Laser-ranging long-baseline
  differential atom interferometers for space},}
  {\protect\JournalTitle{Physical Review A}} \textbf{92}, 063613 (2015).

\bibitem{Liu2018In-Orbit}
L.~Liu, D.-S. L{\"{u}}, W.-B. Chen, T.~Li, Q.-Z. Qu, B.~Wang, L.~Li, W.~Ren,
  Z.-R. Dong, J.-B. Zhao, W.-B. Xia, X.~Zhao, J.-W. Ji, M.-F. Ye, Y.-G. Sun,
  Y.-Y. Yao, D.~Song, Z.-G. Liang, S.-J. Hu, D.-H. Yu, X.~Hou, W.~Shi, H.-G.
  Zang, J.-F. Xiang, X.-K. Peng, and Y.-Z. Wang, \enquote{{In-orbit operation
  of an atomic clock based on laser-cooled 87Rb atoms},}
  {\protect\JournalTitle{Nature Communications}} \textbf{9}, 2760 (2018).

\bibitem{Hogan2011Atomic}
J.~M. Hogan, D.~M. Johnson, S.~Dickerson, T.~Kovachy, A.~Sugarbaker, S.-w.
  Chiow, P.~W. Graham, M.~A. Kasevich, B.~Saif, S.~Rajendran, P.~Bouyer, B.~D.
  Seery, L.~Feinberg, and R.~Keski-Kuha, \enquote{{An atomic gravitational wave
  interferometric sensor in low earth orbit (AGIS-LEO)},}
  {\protect\JournalTitle{General Relativity and Gravitation}} \textbf{43},
  1953--2009 (2011).

\bibitem{Gehler2013ESA}
M.~Gehler, L.~Cacciapuoti, A.~Heske, R.~Biesbroek, P.~Waller, and E.~Wille,
  \enquote{{The ESA STE-QUEST mission study - space mission design to test
  Einstein's equivalence principle},} in \emph{AIAA SPACE 2013 Conference and
  Exposition,}  (American Institute of Aeronautics and Astronautics, Reston,
  Virginia, 2013).

\bibitem{Williams2016Quantum}
J.~Williams, S.-w. Chiow, N.~Yu, and H.~M{\"{u}}ller, \enquote{{Quantum test of
  the equivalence principle and space-time aboard the International Space
  Station},} {\protect\JournalTitle{New Journal of Physics}} \textbf{18},
  025018 (2016).

\bibitem{Hogan2016Atom}
J.~M. Hogan and M.~A. Kasevich, \enquote{{Atom-interferometric
  gravitational-wave detection using heterodyne laser links},}
  {\protect\JournalTitle{Physical Review A}} \textbf{94}, 033632 (2016).

\bibitem{Tino2019SAGE}
G.~M. Tino, A.~Bassi, G.~Bianco, K.~Bongs, P.~Bouyer, L.~Cacciapuoti,
  S.~Capozziello, X.~Chen, M.~L. Chiofalo, A.~Derevianko, W.~Ertmer,
  N.~Gaaloul, P.~Gill, P.~W. Graham, J.~M. Hogan, L.~Iess, M.~A. Kasevich,
  H.~Katori, C.~Klempt, X.~Lu, L.-S. Ma, H.~M{\"{u}}ller, N.~R. Newbury, C.~W.
  Oates, A.~Peters, N.~Poli, E.~M. Rasel, G.~Rosi, A.~Roura, C.~Salomon,
  S.~Schiller, W.~Schleich, D.~Schlippert, F.~Schreck, C.~Schubert,
  F.~Sorrentino, U.~Sterr, J.~W. Thomsen, G.~Vallone, F.~Vetrano, P.~Villoresi,
  W.~von Klitzing, D.~Wilkowski, P.~Wolf, J.~Ye, N.~Yu, and M.~Zhan,
  \enquote{{SAGE: A proposal for a space atomic gravity explorer},}
  {\protect\JournalTitle{The European Physical Journal D}} \textbf{73}, 228
  (2019).

\bibitem{Schuldt2015Design}
T.~Schuldt, C.~Schubert, M.~Krutzik, L.~G. Bote, N.~Gaaloul, J.~Hartwig,
  H.~Ahlers, W.~Herr, K.~Posso-Trujillo, J.~Rudolph, S.~Seidel, T.~Wendrich,
  W.~Ertmer, S.~Herrmann, A.~Kubelka-Lange, A.~Milke, B.~Rievers, E.~Rocco,
  A.~Hinton, K.~Bongs, M.~Oswald, M.~Franz, M.~Hauth, A.~Peters, A.~Bawamia,
  A.~Wicht, B.~Battelier, A.~Bertoldi, P.~Bouyer, A.~Landragin, D.~Massonnet,
  T.~L{\'{e}}v{\`{e}}que, A.~Wenzlawski, O.~Hellmig, P.~Windpassinger,
  K.~Sengstock, W.~von Klitzing, C.~Chaloner, D.~Summers, P.~Ireland,
  I.~Mateos, C.~F. Sopuerta, F.~Sorrentino, G.~M. Tino, M.~Williams,
  C.~Trenkel, D.~Gerardi, M.~Chwalla, J.~Burkhardt, U.~Johann, A.~Heske,
  E.~Wille, M.~Gehler, L.~Cacciapuoti, N.~G{\"{u}}rlebeck, C.~Braxmaier, and
  E.~Rasel, \enquote{{Design of a dual species atom interferometer for space},}
  {\protect\JournalTitle{Experimental Astronomy}} \textbf{39}, 167--206 (2015).

\bibitem{Loriani2019Atomic}
S.~Loriani, D.~Schlippert, C.~Schubert, S.~Abend, H.~Ahlers, W.~Ertmer,
  J.~Rudolph, J.~M. Hogan, M.~A. Kasevich, E.~M. Rasel, and N.~Gaaloul,
  \enquote{{Atomic source selection in space-borne gravitational wave
  detection},} {\protect\JournalTitle{New Journal of Physics}} \textbf{21},
  063030 (2019).

\bibitem{Dimopoulos2009Gravitational}
S.~Dimopoulos, P.~W. Graham, J.~M. Hogan, M.~A. Kasevich, and S.~Rajendran,
  \enquote{{Gravitational wave detection with atom interferometry},}
  {\protect\JournalTitle{Physics Letters B}} \textbf{678}, 37--40 (2009).

\bibitem{Foot2005Atomic}
C.~J. Foot, \emph{{Atomic Physics}}, Oxford master series in physics (Oxford
  University Press, 2005).

\bibitem{Lee1996Single}
K.~I. Lee, J.~A. Kim, H.~R. Noh, and W.~Jhe, \enquote{{Single-beam atom trap in
  a pyramidal and conical hollow mirror},} {\protect\JournalTitle{Optics
  Letters}} \textbf{21}, 1177 (1996).

\bibitem{Arlt1998Pyramidal}
J.~Arlt, O.~Marag{\`{o}}, S.~Webster, S.~Hopkins, and C.~Foot, \enquote{{A
  pyramidal magneto-optical trap as a source of slow atoms},}
  {\protect\JournalTitle{Optics Communications}} \textbf{157}, 303--309 (1998).

\bibitem{Williamson1998Magneto}
R.~Williamson, P.~Voytas, R.~Newell, and T.~Walker, \enquote{{A magneto-optical
  trap loaded from a pyramidal funnel},} {\protect\JournalTitle{Optics
  Express}} \textbf{3}, 111 (1998).

\bibitem{Hinton2017Portable}
A.~Hinton, M.~Perea-Ortiz, J.~Winch, J.~Briggs, S.~Freer, D.~Moustoukas,
  S.~Powell-Gill, C.~Squire, A.~Lamb, C.~Rammeloo, B.~Stray, G.~Voulazeris,
  L.~Zhu, A.~Kaushik, Y.-H. Lien, A.~Niggebaum, A.~Rodgers, A.~Stabrawa,
  D.~Boddice, S.~R. Plant, G.~W. Tuckwell, K.~Bongs, N.~Metje, and M.~Holynski,
  \enquote{{A portable magneto-optical trap with prospects for atom
  interferometry in civil engineering},} {\protect\JournalTitle{Philosophical
  Transactions of the Royal Society A: Mathematical, Physical and Engineering
  Sciences}} \textbf{375}, 20160238 (2017).

\bibitem{Slowe2005High}
C.~Slowe, L.~Vernac, and L.~V. Hau, \enquote{{High flux source of cold rubidium
  atoms},} {\protect\JournalTitle{Review of Scientific Instruments}}
  \textbf{76}, 103101 (2005).

\bibitem{AOSense}
\enquote{{Beam-RevC cold atomic beam system},} Tech. rep., AOSense, Inc.
  Available online at aosense.com/product/cold-atomic-beam-system.

\bibitem{Dieckmann1998Two}
K.~Dieckmann, R.~J.~C. Spreeuw, M.~Weidem{\"{u}}ller, and J.~T.~M. Walraven,
  \enquote{{Two-dimensional magneto-optical trap as a source of slow atoms},}
  {\protect\JournalTitle{Physical Review A}} \textbf{58}, 3891--3895 (1998).

\bibitem{Schoser2002Intense}
J.~Schoser, A.~Bat{\"{a}}r, R.~L{\"{o}}w, V.~Schweikhard, A.~Grabowski, Y.~B.
  Ovchinnikov, and T.~Pfau, \enquote{{Intense source of cold Rb atoms from a
  pure two-dimensional magneto-optical trap},} {\protect\JournalTitle{Physical
  Review A}} \textbf{66}, 023410 (2002).

\bibitem{Cren2002Loading}
P.~Cren, C.~Roos, A.~Aclan, J.~Dalibard, and D.~Gu{\'{e}}ry-Odelin,
  \enquote{{Loading of a cold atomic beam into a magnetic guide},}
  {\protect\JournalTitle{The European Physical Journal D - Atomic, Molecular
  and Optical Physics}} \textbf{20}, 107--116 (2002).

\bibitem{Kellogg2012Compact}
J.~R. Kellogg, D.~Schlippert, J.~M. Kohel, R.~J. Thompson, D.~C. Aveline, and
  N.~Yu, \enquote{{A compact high-efficiency cold atom beam source},}
  {\protect\JournalTitle{Applied Physics B}} \textbf{109}, 61--64 (2012).

\bibitem{Park2012Cold}
S.~J. Park, J.~Noh, and J.~Mun, \enquote{{Cold atomic beam from a
  two-dimensional magneto-optical trap with two-color pushing laser beams},}
  {\protect\JournalTitle{Optics Communications}} \textbf{285}, 3950--3954
  (2012).

\bibitem{Chaudhuri2006Realization}
S.~Chaudhuri, S.~Roy, and C.~S. Unnikrishnan, \enquote{{Realization of an
  intense cold Rb atomic beam based on a two-dimensional magneto-optical trap:
  Experiments and comparison with simulations},}
  {\protect\JournalTitle{Physical Review A}} \textbf{74}, 023406 (2006).

\bibitem{Jollenbeck2011Hexapole}
S.~J{\"{o}}llenbeck, J.~Mahnke, R.~Randoll, W.~Ertmer, J.~Arlt, and C.~Klempt,
  \enquote{{Hexapole-compensated magneto-optical trap on a mesoscopic atom
  chip},} {\protect\JournalTitle{Phys. Rev. A}} \textbf{83}, 43406 (2011).

\bibitem{Jollenbeck2012Eine}
S.~J{\"{o}}llenbeck, \enquote{{Eine Quelle Bose-Einstein-kondensierter
  Ensembles auf Basis eines mesokopischen Atomchips},} {Ph.D.} thesis,
  Gottfried Wilhelm Leibniz Universit{\"{a}}t (2012).

\bibitem{ColdQuanta}
\enquote{{Cold atom source cells},} Tech. rep., ColdQuanta. Available online at
  https://coldquanta.com/standard{\_}products/cold-atom-source-cells/.

\bibitem{Wohlleben2001Atom}
W.~Wohlleben, F.~Chevy, K.~Madison, and J.~Dalibard, \enquote{{An atom
  faucet},} {\protect\JournalTitle{The European Physical Journal D}}
  \textbf{15}, 237--244 (2001).

\bibitem{Donley2005Optical}
E.~A. Donley, T.~P. Heavner, and S.~R. Jefferts, \enquote{{Optical molasses
  loaded from a low-velocity intense source of atoms: an atom source for
  improved atomic fountains},} {\protect\JournalTitle{IEEE Transactions on
  Instrumentation and Measurement}} \textbf{54}, 1905--1910 (2005).

\bibitem{Ovchinnikov2005Compact}
Y.~B. Ovchinnikov, \enquote{{Compact magneto-optical sources of slow atoms},}
  {\protect\JournalTitle{Optics Communications}} \textbf{249}, 473--481 (2005).

\bibitem{Camposeo2001Cold}
A.~Camposeo, A.~Piombini, F.~Cervelli, F.~Tantussi, F.~Fuso, and E.~Arimondo,
  \enquote{{A cold cesium atomic beam produced out of a pyramidal funnel},}
  {\protect\JournalTitle{Optics Communications}} \textbf{200}, 231--239 (2001).

\bibitem{Kohel2003Generation}
J.~M. Kohel, J.~Ramirez-Serrano, R.~J. Thompson, L.~Maleki, J.~L. Bliss, and
  K.~G. Libbrecht, \enquote{{Generation of an intense cold-atom beam from a
  pyramidal magneto-optical trap: experiment and simulation},}
  {\protect\JournalTitle{Journal of the Optical Society of America B}}
  \textbf{20}, 1161 (2003).

\bibitem{Lundblad2004Two}
N.~Lundblad, D.~C. Aveline, R.~J. Thompson, J.~M. Kohel, J.~Ramirez-Serrano,
  W.~M. Klipstein, D.~G. Enzer, N.~Yu, and L.~Maleki, \enquote{{Two-species
  cold atomic beam},} {\protect\JournalTitle{Journal of the Optical Society of
  America B}} \textbf{21}, 3 (2004).

\bibitem{Harris2008Magnetic}
M.~L. Harris, P.~Tierney, and S.~L. Cornish, \enquote{{Magnetic trapping of a
  cold Rb-Cs atomic mixture},} {\protect\JournalTitle{Journal of Physics B:
  Atomic, Molecular and Optical Physics}} \textbf{41}, 035303 (2008).

\bibitem{Imhof2017Two-dimensional}
E.~Imhof, B.~K. Stuhl, B.~Kasch, B.~Kroese, S.~E. Olson, and M.~B. Squires,
  \enquote{{Two-dimensional grating magneto-optical trap},}
  {\protect\JournalTitle{Physical Review A}} \textbf{96}, 033636 (2017).

\bibitem{Steck2015Rubidium}
D.~A. Steck, \enquote{{Rubidium 87 D line data},}  (2015). Available online at
  http://steck.us/alkalidata (revision 2.2.1, 21 November 2019).

\bibitem{Steck2019Cesium}
D.~A. Steck, \enquote{{Cesium D line data},}  (2019). Available online at
  http://steck.us/alkalidata (revision 2.2.1, 21 November 2019).

\bibitem{Tierney2009Magnetic}
P.~Tierney, \enquote{{Magnetic trapping of an ultracold 87Rb-133Cs atomic
  mixture},} {Ph.D.} thesis, University of Durham (2009).

\bibitem{Vangeleyn2009Single}
M.~Vangeleyn, P.~F. Griffin, E.~Riis, and A.~S. Arnold, \enquote{{Single-laser,
  one beam, tetrahedral magneto-optical trap},} {\protect\JournalTitle{Optics
  Express}} \textbf{17}, 13601 (2009).

\bibitem{Li2008Manipulation}
T.~Li, \enquote{{Manipulation of cold atoms using an optical one-way barrier},}
  {Ph.D.} thesis, University of Oregon (2008).

\bibitem{Xu2008Realization}
B.~M. Xu, X.~Chen, J.~Wang, and M.~S. Zhan, \enquote{{Realization of a
  single-beam mini magneto-optical trap: A candidate for compact CPT cold
  atom-clocks},} {\protect\JournalTitle{Optics Communications}} \textbf{281},
  5819--5823 (2008).

\bibitem{Harris2008Realisation}
M.~L. Harris, \enquote{{Realisation of a cold mixture of rubidium and
  caesium},} {Ph.D.} thesis, University of Durham (2008).

\bibitem{McCarron2011Quantum}
D.~McCarron, \enquote{{A quantum degenerate mixture of 87Rb and 133Cs},}
  {Ph.D.} thesis, University of Durham (2011).

\bibitem{Harsono2006Dipole}
A.~Harsono, \enquote{{Dipole trapping and manipulation of ultra-cold atoms},}
  {Ph.D.} thesis, University of Oxford (2006).

\bibitem{Sheard2010Magnetic}
B.~T. Sheard, \enquote{{Magnetic transport and Bose-Einstein condensation of
  rubidium atoms},} {Ph.D.} thesis, University of Oxford (2010).

\bibitem{Lu1996Low}
Z.~T. Lu, K.~L. Corwin, M.~J. Renn, M.~H. Anderson, E.~A. Cornell, and C.~E.
  Wieman, \enquote{{Low-velocity intense source of atoms from a magneto-optical
  trap},} {\protect\JournalTitle{Physical Review Letters}} \textbf{77},
  3331--3334 (1996).

\bibitem{Ravenhall2018Compact}
S.~Ravenhall, \enquote{{A compact, high-flux source of cold atoms},} {Ph.D.}
  thesis, University of Oxford (2018).

\bibitem{Rathod2013Continuous}
K.~D. Rathod, A.~K. Singh, and V.~Natarajan, \enquote{{Continuous beam of
  laser-cooled Yb atoms},} {\protect\JournalTitle{EPL (Europhysics Letters)}}
  \textbf{102}, 43001 (2013).

\bibitem{Gibble1992Improved}
K.~E. Gibble, S.~Kasapi, and S.~Chu, \enquote{{Improved magneto-optic trapping
  in a vapor cell},} {\protect\JournalTitle{Optics Letters}} \textbf{17}, 526
  (1992).

\bibitem{Lindquist1992Experimental}
K.~Lindquist, M.~Stephens, and C.~Wieman, \enquote{{Experimental and
  theoretical study of the vapor-cell Zeeman optical trap},}
  {\protect\JournalTitle{Physical Review A}} \textbf{46}, 4082--4090 (1992).

\bibitem{Hoth2013Atom}
G.~W. Hoth, E.~A. Donley, and J.~Kitching, \enquote{{Atom number in
  magneto-optic traps with millimeter scale laser beams},}
  {\protect\JournalTitle{Optics Letters}} \textbf{38}, 661 (2013).

\end{thebibliography}

\end{document}